\documentclass[12pt,preprint]{aastex}

\begin{document}
\shorttitle{ATTEMPT TO VERIFY INTERSTELLAR GLYCINE}
\shortauthors{Snyder et al.}
\slugcomment{\ Accepted by the ApJ for the 1 Feb. 2005 issue}

\title{A RIGOROUS ATTEMPT TO VERIFY INTERSTELLAR GLYCINE}

\author{L. E. Snyder\altaffilmark{1}, F. J. Lovas\altaffilmark{2}, J. M. 
Hollis\altaffilmark{3}, D. N. Friedel\altaffilmark{1}, P. R. Jewell\altaffilmark{4}, 
A. Remijan\altaffilmark{1,3,5}, V. V. Ilyushin\altaffilmark{6}, E. A. 
Alekseev\altaffilmark{6}, and S. F.  Dyubko\altaffilmark{6}}

\altaffiltext{1}{Department of Astronomy, University of Illinois, Urbana, IL
61801\\ email:  snyder@astro.uiuc.edu, friedel@astro.uiuc.edu,
aremijan@pop900.gsfc.nasa.gov}

\altaffiltext{2}{Optical Technology Division, National Institute of Standards
and Technology, Gaithersburg, MD 20899-8441\\ email:  lovas@nist.gov}

\altaffiltext{3}{Earth and Space Data Computing Div., Code 930, NASA's Goddard
Space Flight Center, Greenbelt, MD 20771\\ email:  Jan.M.Hollis@gsfc.nasa.gov}

\altaffiltext{4}{National Radio Astronomy Observatory, Green Bank, WV
24944-0002\\ email:  pjewell@nrao.edu}

\altaffiltext{5}{National Research Council Resident Research Associate}

\altaffiltext{6}{Institute of Radio Astronomy of NASU, Krasnoznamennaya 4, 61002
Kharkow, Ukraine\\ email:  ilyushin@rian.ira.kharkov.ua; alekseev@rian.ira.kharkov.ua}

\begin{abstract}

In 2003, Kuan, Charnley, and co-workers reported the detection of interstellar
glycine (NH$_2$CH$_2$COOH) based on observations of 27 lines in 19 different
spectral bands in one or more of the sources Sgr B2(N-LMH), Orion KL, and W51
e1/e2.  They supported their detection report with rotational temperature
diagrams for all three sources.  In this paper, we present essential criteria 
which can be used in a straightforward analysis technique to confirm the
identity of an interstellar asymmetric rotor such as glycine. We 
use new laboratory measurements of glycine as a basis for applying this analysis 
technique, both to our previously unpublished 12 m telescope data and to the 
previously published SEST data of Nummelin and colleagues. We conclude that key 
lines necessary for an interstellar glycine identification have not yet been found. 
We identify some common molecular candidates that should be examined further 
as more likely carriers of several of the lines reported as glycine. Finally, we 
illustrate that a rotational temperature diagram used without the support of 
correct spectroscopic assignments is not a reliable tool for the 
identification of interstellar molecules.  

\end{abstract}


\section{INTRODUCTION}

Interstellar glycine (NH$_2$CH$_2$COOH) has been unsuccessfully sought using
both single-element telescopes (Brown et al.  1979; Hollis et al.  1980; Snyder
et al.  1983; Beralis et al.  1985; Guelin \& Cernicharo 1989; Combes, Rieu, \&
Wlodarczak 1996; Ceccarelli et al.  2000) and interferometric arrays (Snyder
1997; Hollis et al.  2003a,b).  Recently, Kuan et al.  (2003) reported the
identification of interstellar glycine (NH$_2$CH$_2$COOH) in the direction of
the hot molecular cores Sgr B2(N-LMH), Orion KL, and W51 e1/e2.

We note several unusual aspects about the Kuan et al. (2003) reported glycine 
results.  First, their glycine report was based on the detection of 27 spectral
line features in the sources Orion~KL, W51 e1/e2, and Sgr~B2(N-LMH). They used
the NRAO\footnotemark[7]\footnotetext[7]{The National Radio Astronomy Observatory 
(NRAO) is a facility of the National Science Foundation operated under cooperative 
agreement by Associated Universities, Inc.} 12~m radio telescope
operating in the range of 130 to 242 GHz. The 27 
independent spectral line features detected in 3 different sources might seem 
to suggest reasonable evidence. Certainly the detection of the same 
ensemble of spectral lines in multiple sources is a characteristic of 
interstellar molecules with structural components similar to glycine, such 
as formic acid (HCOOH) and acetic acid (CH$_{3}$COOH)(see, for example, Liu, 
Mehringer, \& Snyder 2001; Remijan et al.  2002). However, each of the three 
glycine sources was observed to contain no more than 13 to 16 of the 27
independent spectral lines features and only 3 of the 27 spectral line features
were found to be common to all 3 sources. Second, 16 of the 27 reported glycine
features are not the four-fold degenerate rotational transitions that should be
the most likely to be detected for R-branch asymmetric rotor transitions.
Third, the glycine rest frequencies used in their observations were based on a 
Hamiltonian calculation which was extrapolated over 100 GHz.  This method can 
yield potentially large uncertainties in the frequency predictions.  Fourth, 
the glycine column densities derived from rotational temperature diagrams 
were surprisingly high. In fact, Orion KL was reported to have the highest 
glycine column density, but the interstellar molecule with structure 
closest to glycine, acetic acid, has never been detected there (Remijan et al.  2003).

In this paper, we investigate the above concerns by starting from new laboratory
measurements of glycine that improve the Hamiltonian fit and thereby result in 
highly reliable interpolated frequency predictions which provide the foundation 
for our discussion.  We list and explain rigorous and essential criteria 
for identifying a new interstellar molecule such as glycine. 
Then we use these criteria on key glycine
spectral lines reported by Kuan et al. (2003) to interpret independent
observational data for Orion KL, W51 e1/e2, and Sgr~B2(N-LMH), and show 
that their key lines fail the tests required for a correct glycine
identification.  In investigating alternative identifications, we find that
the carriers of several important spectral lines
which have been assigned to glycine are more logically assigned to other,
more common species.  Finally, we find that random choices of unidentified 
spectral lines can be used to generate a reasonable glycine rotational 
temperature diagram similar to those published by Kuan et al.  (2003).  

\section{GLYCINE SPECTRAL LINE ANALYSIS AND PREDICTED REST FREQUENCIES}

The microwave spectrum of conformer I of glycine was first recorded by Suenram
\& Lovas (1980) from 82 GHz to 113 GHz in a heated parallel plate Stark cell.
This study provided the rotational analysis and rough dipole moment
determination.  Further measurements of the $^{14}$N quadrupole coupling
hyperfine structure and Stark effect on low J transitions between 16 GHz and
24.5 GHz were subsequently reported by Lovas et al.  (1995).  While these
literature data provided a firm basis for predicting transitions up to about 150
GHz for guiding past interstellar searches, the 2 $\sigma$ (standard deviation)
uncertainties\footnotemark[8]\footnotetext[8]{Unless otherwise noted, all 
uncertainties quoted in this
paper are 2 $\sigma$ or coverage factor k = 2 as described by Taylor \& Kuyatt
(1994).}  ranged from 1.5 MHz at 206 GHz to 2.8 MHz at 240 GHz for the main
interstellar transitions reported by Kuan et al. (2003).  Recently, at Kharkow a
heated quartz absorption cell was used to provide new measurements between 75
GHz and 260 GHz utilizing an automated synthesizer-based spectrometer described by
Ilyushin et al.  (2001).  New glycine search frequencies could then be predicted
by incorporating both the previously reported transitions and the new
measurements into a rotational analysis which employed the Watson A-reduction
Hamiltonian with quartic centrifugal distortion terms included (Watson 1977).
The details of these laboratory measurements and Hamiltonian calculations will
be reported in a future paper.  For this paper, it is important to note that for
the majority of the calculated transitions the uncertainties have been reduced
by a factor of 10 or more from those used by Kuan et al.  (2003). Thus, we are 
starting our analysis with highly accurate glycine rest frequency predictions.

Table 1 summarizes the reported glycine line detections in the direction of the
hot molecular cores Sgr B2(N-LMH), Orion KL, and W51 e1/e2.  The first column
lists a number that Kuan et al. (2003) introduced to represent a particular 
glycine line. For the convenience of the reader, we adopt this shorthand notation 
hereafter referred to as the "glycine line number". The second column
gives the rotational quantum numbers of the associated transitions, and the
third identifies each reported transition as an a-type or a b-type.  The fourth
column lists our newly calculated glycine rest frequencies, and the fifth and
sixth columns give the associated line strengths and upper energy levels.  The
final six columns in Table 1 list the observed LSR velocity and brightness
temperature of each glycine line reported for each source.  We have listed in
boldface the LSR velocity and brightness temperature values for the lines that
were displayed by Kuan et al.  (2003).

\section{ESSENTIAL CRITERIA FOR ESTABLISHING THE IDENTIFICATION OF A NEW
INTERSTELLAR MOLECULE}

The task of establishing the identification of a new interstellar molecule in a
dense interstellar cloud is made more complicated by high spectral line
densities.  For example, at 3 mm wavelength Sgr B2(N-LMH) has a spectral line
density of 6.06 lines per 100 MHz, half of which are unidentified (Friedel et
al.  2004).  This is the highest reported line density in radio astronomy.
Hence, the detection of spectral lines with frequencies close to a new species
is not in itself sufficient evidence for a correct assignment.  Consequently,
there are several criteria to consider in securing the correct identification of
a new interstellar molecule such as glycine.

{\it Rest Frequencies}:  The most important criterion for establishing the
identification of a new interstellar molecule is that the rest frequencies must
be established to a high degree of accuracy.  Preferably, the spectral line
astronomical rest frequencies to be used in the interstellar search should have
been directly measured in the laboratory.  If such direct measurements are not
available, interpolation fitting of the available laboratory data from other
frequencies should allow construction of a high precision Hamiltonian model
which can be used to predict search frequencies with uncertainties on the order
of 1 part in 10 million.

{\it Frequency Agreement}:  There must be frequency agreement among all detected
transitions.  This means that an accurate astronomical rest frequency of the
assigned transition must be in reasonable agreement with the frequency
corresponding to the LSR velocity of the source.  Variations in the LSR velocity
of a given molecular species can occur when regions with different kinematics
contribute to the emission or absorption spectrum.  However, it is evident that
emission lines from a given molecular species which emanate from a region with
no large kinematic variance will have a well-defined velocity field.  Examples
are the VLA measurements of the kinematics of methyl formate (HCOOCH$_3$)
toward OMC-1 by Hollis et al. (2003a) and ethyl cyanide (CH$_3$CH$_2$CN)
toward Sgr B2(N-LMH) by Hollis et al.
(2003b).  If there is a source velocity gradient, as established from observed
transitions of known molecules, it cannot be a random function of transition
frequency or energy level.

Furthermore, it is not uncommon to have an interstellar spectral line from
one species blend with another line of either the same or another species.
In order to correctly assign or confirm a line identification, a common
standard is that the lines should be at least resolved by the Rayleigh
criterion.  That is, the minimum distance between two spectral lines
in frequency space must be such that the maximum intensity of one
line falls on the first null of the other (see, for example, Jenkins \&
White 1957; Sommerfeld 1964). This is approximately equivalent to requiring 
that overlapping lines be separated by their full line width at half maximum 
intensity (assuming identical line profiles).  In practice, a more stringent
criterion based on the signal-to-noise ratio is often needed whereby two
overlapping lines can be considered resolved if they are at least 
separated at half-maximum intensity of the weakest line.

{\it Beam Dilution}:  Suppose one uses a single-element radio telescope with
circular aperture of diameter D (such as the NRAO 12 m) to observe separate 
transitions of some molecule at 50 GHz, 100 GHz, and 150 GHz.  The full 
half-power beam width
(HPBW) of the telescope is approximately given by 1.22~$\lambda$/D rad (Born \&
Wolf 1980; see also Rohlfs \& Wilson 2000), so in this case the telescope HPBW
changes by a factor of 3 over the range of observations. As we will discuss,
beam dilution scales with source size squared and beam size squared. To examine
the required correction for beam dilution, we can start with $<N_T>$, the
beam-averaged molecular column density determined from the integrated intensity
of a rotation transition observed by a single-element radio telescope.  In
autocorrelation mode, with the assumptions of LTE and low optical
depth\footnotemark[9]\footnotetext[9]{The corrections to apply when the 
optical depth is not low and when the excitation is subthermal have been 
discussed by in detail by Goldsmith \& Langer (1999) and are mentioned in 
Appendix A. The aspects of maser excitation have been examined by 
Reid \& Moran (1988). Because we are following
the assumptions of Kuan et al. (2003) that their reported glycine lines are 
optically thin and in LTE, we will confine our discussion to this particular 
case.} (see, for example, Snyder et al. 2001; Dickens et al. 1997), $<N_T>$ 
is given by

$$ <N_T> = {1.67~W_{\rm T}~Z~e^{E_{\rm u}/T_{\rm rot}}\over S \mu^2 \nu}
\times 10^{14}~ {\rm cm}^{-2}.  \eqno(1) $$

\noindent In equation (1), $W_{\rm T} = \int T_R~dv$ in K km s$^{-1}$, 
where T$_R$ is the radiation temperature of the source and $dv$ is the 
FWHM line width. T$_R$ is related to the measured antenna temperature 
$T_R^*$ (corrected for atmospheric attenuation, rear spillover, blockage, 
ohmic losses, and forward spillover) by T$_R$ = 
$T_R^*$/$\eta _C$, 
where $\eta _C$ is the source-beam coupling efficiency (Kutner \& Ulich 
1981). Other terms in equation (1) are $Z$, the rotational partition function,
E$_{\rm u}$, the upper rotational energy level, T$_{\rm rot}$, the 
rotational temperature, S, the line strength, $\mu^2$, the square of 
the dipole moment in Debye$^2$, and $\nu$, the frequency in GHz. If 
the source is comparable to or smaller than the main beam, 
$\eta _C$ = $\eta _M^*$~B, where $\eta _M^*$ is the corrected main 
beam efficiency and is the fraction of power in the main beam relative 
to the power in the main beam plus error beam (Jewell 1990). B is the 
geometric coupling efficiency between the main beam and the source, and 
is often called the beam filling factor. Note that $\eta _M^*$ is related 
to the phase errors of the reflector surface and 
is thus a function of frequency. For example, for the NRAO 12 m, 
$\eta _M^*$ = 0.83 at 113 GHz and 0.45 at 240 GHz. The above formalism is 
applicable to sources with angular extent smaller than or on the order of 
the main beam. Highly extended sources may couple with some or all 
of the error beam and require a different source-beam convolution.

For a circular Gaussian telescope beam size $\Theta_b$ centered on the 
peak of a circular Gaussian source of size $\Theta_s$, the resulting 
convolution yields a beam filling factor B given by (e.g., see equation 
(28) of Ulich \& Haas 1976)

$$ B = {\Theta_s^2\over \Theta_b^2 + \Theta_s^2}.  \eqno(2) $$

\noindent If $\Theta_s$ $\gg$ $\Theta_b$, B $\sim$ 1.  When
$\Theta_s$ $\sim$ $\Theta_b$, B $\sim$ 0.5, but if $\Theta_s$ $\ll$ $\Theta_b$,
B $\sim$ $\frac{\Theta_s^2}{\Theta_b^2}$.

{\it Relative Intensities}:  Once several molecular transition assignments 
have been made, their relative intensities must be tested for consistency.
For a source comparable to or smaller in angular size than the main beam 
of the telescope, the relative intensity of two single lines is determined 
by the corrected beam efficiencies, beam filling factors, dipole moments, 
line strengths, frequencies, energy levels, and rotational temperature 
according to equation (1) as

$$ \frac{T_R^*(i)}{T_R^*(j)} = \left( {\frac{\eta _M^*(i)}{\eta 
_M^*(j)}} \right)\left( \frac{B_i}{B_j} \right) 
\left(\frac{\nu_i}{\nu_j} \right)\left(\frac{S_i}{S_j} \right) 
\left( \frac{\mu^{2}_i}{\mu^{2}_j} \right)
e^{-\frac{E_{ui}-E_{uj}}{T_{rot}}}.  \eqno(3) $$

\noindent Conformer I glycine has an a-type dipole moment, $\mu_a$ = 
0.911(6) D (3.039(20) x 10$^{-30}$ C m), and a b-type, $\mu_b$ =
0.697(10) D (2.325(34) x 10$^{-30}$ C m) (Lovas et al. 1995).
Therefore, for a given frequency, the product S$\mu$$^2$ dictates that 
a-type transitions are stronger than b-type when S values are comparable.

A typical property of prolate asymmetric rotors such as glycine is that two
low K$_{-1}$ a-type and two b-type transitions become degenerate as the 
transition frequencies increase. In this four-fold degeneracy, all four transitions
have the same rest frequency and energy levels, but different line strengths 
for the a-type and the b-type. This effect begins at $\sim$ 107 GHz for 
glycine and it is important because the lines with this degeneracy are the most likely 
to be detected. The total intensity of a four-fold degenerate line at frequency
$\nu_i$ is given by the sum of the four individual unresolved components,
$\Bigg({\displaystyle \sum_{k=1}^{4}T_R^*(k)\Bigg)_i}$, and the ratio of the 
{\it ith} degenerate line to the {\it jth} degenerate line is given by 
equation (1) as

$$\frac{\Bigg({\displaystyle \sum_{k=1}^{4} 
T_R^*(k)}\Bigg)_i}{\Bigg({\displaystyle \sum_{k=1}^{4} T_R^*(k)}\Bigg)_j} = 
\left( {\frac{\eta _M^*(i)}{\eta
_M^*(j)}} \right)\left(
\frac{B_i}{B_j} \right) \left( \frac{\nu_i}{\nu_j} \right) \left(
\frac{S_{ai}\mu_a^{2}+S_{bi}\mu_b^{2}} {S_{aj}\mu_a^{2}+ S_{bj}\mu_b^{2}}
\right) e^{-\frac{E_{ui}-E_{uj}}{T_{rot}}}.  \eqno(4) $$

Similar expressions can be derived for comparing the intensity of a single
glycine line to either a two-fold or a four-fold degenerate glycine line.

{\it Confirming Transitions}:  Once a candidate line is assigned, the assumption
of optical thinness under LTE conditions allows predictions of intensities of
additional confirming transitions unless line self-absorption, maser activity,
or some other mitigating effect is evident.  Hence, a key test of the
correctness of the assignment of a transition is that any other transitions
connected by favorable transition probabilities must also be present if the
relative intensity predictions lead to detectable signal levels.

The maximum line strength for a-type asymmetric rotor R-branch transitions 
of a molecule like glycine occurs when S approaches J$'$, the upper level 
value of J.  For a-type transitions, this means that the strongest 
transitions in a series will be those for which 
J$'$$_{\rm K'_{-1},\rm K'_{1}}$ - J$_{\rm K_{-1},\rm K_{1}}$ =
J$'$$_{0, \rm J'}$ - J$'$-1$_{0,\rm J'-1}$ or 
J$'$$_{1,\rm J'}$ - J$'$-1$_{1,\rm J'-1}$. 
The next strongest in this series will be 
J$'$$_{1, \rm J'-1}$ - J$'$-1$_{1,\rm J'-2}$ or 
J$'$$_{2,\rm J'-1}$ - J$'$-1$_{2,\rm J'-2}$.
For b-type transitions, the maximum line strength occurs
when S approaches J$'$-1.  Here the strongest transitions will be those 
for which
J$'$$_{0,\rm J'}$ - J$'$-1$_{1,\rm J'-1}$ or 
J$'$$_{1,\rm J'}$ - J$'$-1$_{0, \rm J'-1}$.
The next strongest will be 
J$'$$_{1,\rm J'-1}$ - J$'$-1$_{2,\rm J'-2}$ or
J$'$$_{2,\rm J'-1}$ -J$'$-1$_{1, \rm J'-2}$.
Note that a-type transitions follow radiative selection
rules where the parity change of the quantum numbers K$_{-1}$,K$_{1}$ in a
transition is e,e $\leftrightarrow$ e,o (even, even $\leftrightarrow$ even, 
odd) or o,o $\leftrightarrow$ o,e while the b-type transitions obey 
e,e $\leftrightarrow$ o,o or e,o $\leftrightarrow$ o,e (see, for example, 
Townes \& Schawlow 1975).

\section{AN EXAMINATION OF THE DATA}

Kuan et al. (2003) state that their 27 glycine lines (Table 1) are detected 
in one or more hot molecular cores (HMCs). We note that HMCs are compact 
objects where interferometric measurements have shown that structurally
similar but simpler large molecules are concentrated with diameters $<$ 10$"$. 
For example, the HMC of one of their sources, Sgr~B2(N-LMH), is 
$\sim$ 4$"$ as measured in ethyl cyanide (CH$_3$CH$_2$CN) and vinyl cyanide 
(CH$_2$CHCN) by Liu \& Snyder (1999). On the other hand, Kuan et al. (2003) 
assume that their HMC glycine lines are optically thin and in LTE and do not 
correct for beam dilution in their rotational temperature diagrams- a tacit 
assumption that their glycine sources are extended enough to at least fill 
their beam. In the case of Orion KL, they speculated that the glycine 
emission is extended because the rotational temperature diagram has a "tight 
fit" and, in addition, perhaps the emission is not perfectly centered at 
the nominal compact ridge position. They used this scenario to reject the 
Orion glycine negative results found by Combes et al. (1996) with the 
IRAM 30~m telescope because the narrower 30~m beam might have missed a 
significant portion of the glycine emission observed by the wider 
NRAO~12~m beam.

We note that equation (1) of Kuan et al.(2003) is basically the 
same as equation (23) of Goldsmith \& Langer (1999) if the gas is assumed 
to be optically thin in LTE. The resulting column density is called a 
beam-averaged total column density. However, equation (38) of Goldsmith 
\& Langer (1999) shows that equation (1) actually is based
on the further assumption that each glycine source fills the beam at each 
of the reported glycine frequencies. Otherwise, the derived glycine 
beam-averaged total column densities and rotational excitation 
temperatures have little physical meaning; they just become numerical 
parameters of a least-squares fit with little predictive value. 
How can the  condition that the reported glycine sources fill each beam be 
satisfied? Kuan et al. (2003) state that their half-power beamwidths 
are $\sim$ 30$"$, 45$"$, and 60$"$ at 1.3, 2, and 3 mm, respectively. 
Table 1 of  Kuan et al. (2003) lists the 2 mm lines (starting with 
glycine line 1)  as the lowest frequency glycine lines that they claim. 
Thus these lines would determine the smallest extent ($\sim$ 45")  
that the reported glycine could have in equation (1) of  Kuan et al. (2003) 
without suffering from beam dilution. Consequently, in our analysis we used 
45$"$ as our upper limit for our trial source size. 
 
In the following, we will apply the criteria from \S 3 to examine available
observational data for the possibility that the spectral lines assigned to 
glycine have been misidentified.  In order to address the ambiguities 
introduced by the glycine source sizes, we will assume two cases: (1) the
glycine emission sources are HMCs with core diameters typically 
$<$ 10$"$; or (2) glycine fills the beam of the 12~m telescope and hence 
is extended over at least 45$"$ in each source. For our analysis we
will use our newly determined astronomical rest frequencies listed in Table 1.

\subsection{VLA Searches for Glycine in Compact Sources:  Orion and
Sgr~B2(N-LMH)}

In 2001, Hollis et al.  (2003b) conducted a deep Q-band 
($\lambda$ $\sim$7 mm) search with the Very Large Array (VLA) toward OMC-1 
in four rotational transitions of conformer I glycine. The J2000.0 phase 
center for their observations was $\alpha$ = 5$^h$35$^m$14$^{s}$.25 and 
$\delta$ = -5$^\circ$22$'$35$"$.5, which is midway between the Orion 
hot core source and the peak formic acid position of Liu et al.  (2002).  
The Kuan et al.  (2003) pointing position precessed from B1950.0 to 
J2000.0 was $\alpha$ = 5$^h$35$^m$14$^s$.48 and 
$\delta$ = -5$^\circ$22$'$36$''$.6 toward Orion KL. The average 12 m 
telescope half-power beam width (HPBW) for the Orion
observations of Kuan et al.  (2003) was $\sim$ 41$"$ and the VLA primary beam
$\sim$ 45$"$, so the difference in pointing positions is not significant.  The
best VLA upper limits of Hollis et al.  (2003b) were for the 7$_{0,7}$ -
6$_{0,6}$ transition at 47,753.841(28) MHz with S = 6.851, E$_{\rm u}$
= 8.657 K, and synthesized telescope beam size $\Theta_b$ = 6$''$:  
$<N_T>$ $<$ 8.0
$\times~10^{14}$ cm$^{-2}$ for $T_{\rm rot}$ = 43 K; and $<N_T>$ $<$ 25.2
$\times~10^{14}$ cm$^{-2}$ for $T_{\rm rot}$ = 100 K.  Kuan et al.  (2003) found
$<N_T>$ = 4.37 $\times~10^{14}$ cm$^{-2}$ for $T_{\rm rot}$ = 141 K, which is
below the upper limits of Hollis et al.  (2003b) only if there is no significant
beam dilution in the 12 m results for Orion KL.  However, if the glycine source
size is 6$"$ in diameter (suggested by Hollis et al.  2003b), then the 12 m
glycine column density $<N_T>$ becomes 2.08 $\times~10^{16}$ cm$^{-2}$, which is
far above the VLA upper limits.

In 2003, Hollis et al. (2003a) repeated their VLA Q-band glycine search 
toward Sgr~B2(N-LMH).  Again the results were negative, and the best upper
limits were for the 7$_{0,7}$ - 6$_{0,6}$ transition observed with synthesized
beam 1.$''$5$\times$1.$''$4 in the direction of the K2 position.  They reached
an upper limit of $<N_T>$ $<$ 1.4 $\times~10^{17}$ cm$^{-2}$ for $T_{\rm rot}$ =
170 K.  In this same source, Kuan et al.  (2003) reported $<N_T>$ = 4.16
$\times~10^{14}$ cm$^{-2}$ for $T_{\rm rot}$ = 76 K, from observations with an
average HPBW of 38$''$.  If the glycine source size is comparable to the VLA
synthesized beam, then the 12 m glycine column density $<N_T>$ becomes 2.67
$\times~10^{17}$ cm$^{-2}$, which is again above the VLA upper limits.

In summary, the VLA glycine negative results (Hollis et al. 2003b; 2003a)
contradict the glycine detection reports only if the Orion glycine sources in
Orion and Sgr~B2(N-LMH) are compact. However, as discussed previously, 
Kuan et al. (2003) assumed that their glycine emission sources are extended
and at least 45$"$ in diameter. In the following sections, we will examine 
this assumption.

\subsection{Search for Glycine in Extended Sources:  Orion KL}

A direct way to examine the assumption of Kuan et al. (2003) that their reported
glycine line sources are extended is to use their reported results to generate
predicted intensities for other glycine lines that they did not observe.  These
predictions can be checked against other single-element telescope observations
that would be sensitive to glycine emission from extended sources. In 
particular,
we observed interstellar acetone [(CH$_3$)$_2$CO] in 1995 March with the NRAO
12~m radio telescope toward Sgr~B2(N) (Snyder et al.  2002).  During Orion time,
we searched for the nearly four-fold degenerate J = 19-18 glycine transitions
around 113,336 MHz.  Our Orion KL pointing position was essentially identical to
that used by Kuan et al.  (2003):  exactly the same right ascension and within
9$"$ in declination.  The NRAO 3 mm single-sideband SIS receiver had sideband
rejection $\geq$ 25 db and effective system temperature (referenced to above the
atmosphere and including rear and forward spillover efficiencies) of $\sim$ 
300~K.  Chopper calibration corrected for atmospheric extinction and telescope
losses and the resultant data are on the T$_R^*$ temperature scale (Kutner and
Ulich 1981).  Data were taken while position switching 30$'$ in azimuth, while
Kuan et al.  (2003) switched by 20$'$.  The half
power beam width (FWHM) was $\sim$56$''$ at 113.3 GHz.  The spectrometer
consisted of the NRAO Hybrid Spectrometer with two 256 channel filter banks used
as a backup.  The Hybrid Spectrometer was operated with two polarization IFs in
parallel, with 300 MHz bandwidth and 768 Hanning-smoothed channels per IF,
giving an effective spectral resolution of 0.781 MHz.  The filter banks were
operated with two IFs in parallel, with 250 kHz resolution in one bank and 500
kHz in the other.  Therefore, this system was exactly the same as that used by
Kuan et al.  (2003) except that they chose to use wider filters for their
observations.  The NRAO 12~m data were reduced using the NRAO data reduction
package UniPOPS (Salter, Maddalena, \& Garwood 1995)
\footnotemark[10]\footnotetext[10]{UniPOPS information is available at
 http://www.gb.nrao.edu/$\sim$rmaddale/140ft/unipops/unipops\_toc.html.}. 

The Kuan et al.  (2003) pointing position toward Orion KL was $\alpha$(B1950.0)
= 5$^h$32$^m$47$^s$.0 and $\delta$(B1950.0) = -5$^\circ$24$'$30$''$.0.  Our
pointing position was $\alpha$(B1950.0) = 5$^h$32$^m$47$^s$.0 and
$\delta$(B1950.0) = -5$^\circ$24$'$21$''$.0, which in our $\sim$ 56$"$ beam was
only a negligible 9$"$ difference.  Table 2 summarizes our
Orion KL search frequencies for the J = 19-18 glycine transitions and for the
test lines used to check the operation of the system.  The first column lists
rest frequencies; the second and third list molecular identifications (when
known) and rotational quantum numbers; the fourth, fifth, and sixth columns give
the transition types, line strengths, and upper energy levels; and the next two
columns give intensities and line widths for Orion KL.
Figure 1 shows the Orion KL spectra
from 113,323 to 113,354 MHz (centered at 113,339 MHz) observed with the Hybrid
Spectrometer on the NRAO 12~m.  The spectral positions of the negative results
for the nearly four-fold degenerate J = 19-18 glycine lines (listed in Table 2)
are marked by the four vertical lines. The unidentified line U113226 is 
on the left, and the 6$_1$-6$_0$ E$_1$ transition of CH$_3$OD
is on the right; these two lines served as system checks.  

Kuan et al.(2003) reported the detection of 15 optically thin glycine lines 
in the direction of Orion KL. If they were glycine, lines 1, 5, 6, 8, 9, 
14, 15, 16, 17, 18, and 25 in Table 1 would be single transitions.  Lines 
11, 21, 24, and 26 would be four-fold degenerate. Of these 15 lines, none 
contained connecting confirming transitions, which would be a key test of the 
correctness of the glycine assignment as discussed in \S 3. Among these, 
we note that lines 5, 14, 15, 16, and 26 marginally meet the Rayleigh 
criterion for spectral resolution. By applying the discussion in \S 3 and 
adapting equation 4 to {\it each} of the reported Orion glycine lines, we 
can use each reported line to make an independent prediction of the 
intensity of the nearly four-fold degenerate J = 19-18 glycine lines that 
we did not detect, as shown in Figure 1. 
Figure 2 shows the range of predicted intensities that we should have
observed with the NRAO 12~m telescope for each of the 15 of the reported 
Orion glycine transitions emanating from a 45$"$ extended source. 
For each prediction, three different points
are labeled to denote the Kuan et al. (2003) rotational temperature and its 
uncertainties, T$_{\rm rot}$ = 141$^{+76}_{-37}$ K. The 
uncertainties for each rotational temperature point
are based on the rms noise level given for each reported 
glycine line by Kuan et al. (2003) (see Table 1).
The dotted line in Figure 2 shows the 3.7 mK noise level from Figure 1.
For example, Orion line 21 has a peak intensity T$_{R}$* = 230$\pm$ 10.0 mK 
as observed with a 30" beam.  In Table 1, it is assigned as a degenerate 
quartet of two a-type and two b-type glycine lines at 206,468 MHz: 
35$_{0,35}$-34$_{0,34}$; 35$_{1,35}$-34$_{1,34}$; 35$_{0,35}$-34$_{1,34}$; 
and 35$_{1,35}$-34$_{0,34}$. If line 21 is an optically thin glycine line 
in LTE with a rotational temperature of 141 K, then the nearly blended 
J = 19 - 18 quartet would appear in Figure 1 with peak intensity 
T$_{R}$* $\sim$ 143 mK $\pm$ 7 mK for an extended source size of 
$\Theta_s$ = 45$"$. 
All of the predictions plotted in Figure 2 are above the 3.7 mK noise level
from Figure 1. Hence, if the Orion reported glycine detection were correct,
we should have detected
a blended line with intensity 5 mK $\leq$ T$_R^*$ $\leq$ 205 mK for the J = 19
- 18 glycine quartet.  Since we did not, we conclude that glycine has not been
detected in an extended source in Orion KL.  In \S 4.1, we ruled out the
detection of glycine in a compact source in Orion.  Consequently, we conclude
that the presence of interstellar glycine has not been verified in Orion~KL.

\subsection{Search for Glycine in Extended Sources:  W51}

During the 1995 March observations with the NRAO 12~m, we also searched for the
four-fold nearly degenerate J = 19 - 18 glycine transitions in the direction of
W51.  The Kuan et al.  (2003) pointing position toward W51~e1/e2 was
$\alpha$(B1950.0) = 19$^h$21$^m$26$^s$.3 and $\delta$(B1950.0) =
+14$^\circ$24$'$39$''$.0.  Our W51 pointing position was $\alpha$(B1950.0) =
19$^h$21$^m$26$^s$.3 and $\delta$(B1950.0) = +14$^\circ$24$'$43$''$.0, which in
our $\sim$ 56$"$ beam was only a negligible 4$"$ difference.  As with the
Orion~KL observations, our W51 data were taken while position switching 30$'$
in azimuth, while Kuan et al.  (2003) switched 20$'$.  Again, the system
was exactly the same as that used by Kuan et al.  (2003) except that they used
wider filters. In Table 2, the last two columns give intensities and line widths 
for the W51 search frequencies for the J = 19-18 glycine transitions and for 
the test lines used to check the operation of the system; they are the same 
transitions used for the Orion KL search. Figure 3 shows the W51 spectra 
from 113,323 to 113,354 MHz (centered at 113,339 MHz) observed with the Hybrid 
Spectrometer on the NRAO 12~m.  As in the case of Orion KL, the spectral 
positions of the negative results for the nearly four-fold degenerate J = 
19-18 glycine lines (listed in Table 2) are marked by the four vertical lines 
centered at 113,336 MHz. As in Figure 1, the system checks are the unidentified 
line U113226 on the left, and the weak 6$_1$-6$_0$ E$_1$ transition of CH$_3$OD 
on the right.

Kuan et al. (2003) reported the detection of 16 optically thin glycine lines in
the direction of W51.  If these lines were glycine, lines 3, 4, 5, 7, 8, 9, 
13, 15, and 25 in Table 1 would be single transitions.  Line 12 would be 
two-fold degenerate and lines 19, 20, 21, 24, 26, and 27 would be four-fold 
degenerate. Of these 16 lines, none contained connecting confirming transitions, 
which would be a key test of the correctness of the glycine assignment as 
discussed in \S 3. Among the displayed spectra, line 27 does not 
meet the Rayleigh criterion for spectral resolution. As with the Orion KL 
data, we can use the intensity of each of the reported glycine lines to 
independently predict the intensity of the nearly four-fold 
degenerate J = 19-18 glycine lines at 113,366 MHz that we did not 
detect, as shown in Figure 3. Figure 4 shows the range of predicted 
intensities that we should have observed with the NRAO 12~m telescope for 
each of the 16 reported W51 glycine transitions emanating from 
a 45$"$ extended source. As in Figure 2, three different points
are labeled for each prediction to denote the Kuan et al. (2003) rotational 
temperature and its uncertainties, T$_{\rm rot}$ = 121$^{+71}_{-32}$ K. 
The uncertainties for each rotational temperature point are based on 
the rms noise level given for 
each reported glycine line by Kuan et al. (2003) (see Table 1). The dotted 
line in Figure 4 shows the 6.4 mK noise level from Figure 3. All of the 
predictions plotted in Figure 4, except those from line 19, are above the 
this noise level. Hence, the intensities of 15 out of 16 glycine lines 
predict that we should have detected a blended J = 19 - 18 glycine line 
quartet with intensity 10 mK $\leq$ T$_R^*$ $\leq$ 105 mK if the reported 
W51 glycine detection is correct. Since we did not, we conclude that 
the presence of interstellar glycine has not been verified in W51.

\subsection{Search for Glycine in Extended Sources:  Sgr~B2(N-LMH)}

Kuan et al. (2003) assigned glycine transitions to 13 emission lines in the
direction of Sgr~B2(N-LMH) at $\alpha$(B1950.0) = 17$^h$44$^m$10$^s$.20 and
$\delta$(B1950.0) = -28$^\circ$21$'$15$''$.0.  Their rest frequencies were
calculated with respect to V$_{\rm LSR}$ = 64 km~s$^{-1}$.  Their emission lines
1, 2, 6, and 7 in Table 1 were assigned to single glycine transitions.  Line 10
would have a five-fold degeneracy and line 12 a two-fold if they were glycine.  
Lines 19, 20, 21, 22, 23, 24, and 26 would be four-fold degenerate.  
Of these 13 lines, only lines 21 and 22
would contain connecting confirming transitions, a key test of the correctness
of the assignment as discussed in \S 3.  Only lines 19, 20, 21, 23, and 26 were
displayed by Kuan et al.  (2003); of these, lines 19, 21, and 23 marginally
meet the Rayleigh criterion for spectral resolution.  Fortunately, Nummelin et
al.  (1998) surveyed Sgr~B2(N) in the direction $\alpha$(B1950.0) =
17$^h$44$^m$10$^s$.10 and $\delta$(B1950.0) = -28$^\circ$21$'$17$''$.0 between
218.3 and 263.55 GHz with the 15 m Swedish ESO-Submillimetre Telescope (SEST).
Their position coincided with the Sgr~B2(N-LMH) position used by Kuan et al.
(2003). Their beam width was $\sim$ 20$"$ (adequate for extended sources), and
their spectral resolution was $\sim$ 1.4 MHz, which corresponds to 
$\sim$ 1.8 km~s$^{-1}$ at 230 GHz. 
Their frequency range covered several 
important glycine transitions, which we will discuss. Rest frequencies 
were calculated with respect to V$_{\rm 
LSR}$ = 62 km~s$^{-1}$.

By applying the discussion in \S 3 to predict the expected glycine 
intensities, it is straightforward to use the Nummelin et al. (1998) data 
to examine the Sgr~B2(N-LMH) glycine assignments of Kuan et al. (2003).
If lines 26 and 27 were glycine, for any reasonable temperature their line 
strengths would make them the the two strongest glycine lines listed in 
Table 1.  
Line 26 was observed by Kuan et al. (2003) with the NRAO 12m telescope to 
have T$_R^*$ = 94 mK.  It was assigned to a degenerate glycine quartet 
consisting of two a-type and two b-type transitions with J = 40 - 39, 
K$_{-1}$ = 1 or 2 at 240,899.5 MHz. Nummelin et al. (1998) detected this 
line with SEST; it has a main beam
brightness temperature T$_{mb}$ = 300~mK. Kuan et al. (2003) assigned line
27 to a degenerate glycine quartet with J = 41 - 40, K$_{-1}$ = 0 or 1 at
241,373.3 MHz.  While line 27 would be expected to be slightly stronger 
than line 26, it was reported by Kuan et al. (2003) to be masked by 
unidentified interlopers in their Sgr~B2(N-LMH) 12m data and the spectrum
was not displayed. However, this interloper masking argument is not 
supported by the data of Nummelin et al. (1998), which show not only 
that the line 27 frequency falls in a relatively clear spectral region 
in Sgr~B2(N-LMH), but also that line 27 was not detected (as we will discuss).

Nummelin et al. (1998) reported all of their data in units of K on the 
main beam brightness temperature (T$_{mb}$) scale. Hence, $\eta _M^*$, the 
main beam efficiency has already been taken into account, and equation 4
may be used for intensity predictions with T$_{mb}$ substituted for
$T_R^*$/$\eta _M^*$. To conduct our analysis, we used the T$_{mb}$ = 300 mK 
intensity (with an estimated 50 mK rms noise level uncertainty) of line 
26 from Nummelin et al. (2003) as a basis for predicting the 
intensities of those K$_{-1}$ = 1 or 2 degenerate glycine quartet lines 
connected to line 26 via two a-type and two b-type transitions.  By 
confining our predictions to the Nummelin et al. (1998) data set, we 
eliminated any problems caused by calibration differences between the 
NRAO~12~m and SEST telescopes. Table 3 lists all of the K$_{-1}$ = 1 or 2 
degenerate glycine lines that fall in the 218.3 - 263.55 GHz range of the 
Nummelin et al. (1998) survey that would be radiatively connected to 
line 26 if its assignment to the degenerate glycine quartet of two a-type 
and two b-type transitions with J = 40 - 39, K$_{-1}$ = 1 or 2, were correct.  
The first column lists the individual quantum numbers of each degenerate 
frequency group, the second lists the transition type, and the third lists 
the rest frequency of each degenerate glycine line. Note that the line quartets
are arranged in descending order of frequency.  The top transition in each
degenerate quartet feeds directly into the top transition in the next lower
frequency degenerate quartet, the second transition in each degenerate quartet
feeds directly into the second transition in the next lower degenerate quartet,
etc.  The fourth column in Table 3 gives the line strengths and the fifth the
upper state energy levels.  The next column gives the predicted values of
T$_{mb}$ for each line if the assignment of line 26 to the J = 40 - 39, 
K$_{-1}$ = 1 or 2, degenerate glycine transition at 240,899.5 MHz were correct.
These predicted T$_{mb}$ values (as well as the values for the upper and 
lower error bars) were calculated for a gas rotational temperature of 75~K 
with an uncertainty ranging from 59~K to 104~K (Kuan et al.  2003) in addition 
to the 50 mK rms noise level uncertainty for line 26. The next column lists what was 
observed by Nummelin et al. (1998) at each frequency: the peak T$_{mb}$ value 
for any clearly resolved line; the upper limit to the noise level if no line 
is present; or the notation ... indicating that the line is masked by the 
wings of an interloper line. The final column in Table 3 lists relevant 
comments, if any, for each observed frequency. Table 4 follows the format of 
Table 3 for the stronger K$_{-1}$ = 0 or 1 degenerate glycine lines that fall 
in the same range of the Nummelin et al. (1998) survey. Again, the predicted 
intensities in Table 4 are based on the observed intensity of 
T$_{mb}$ = 300~mK for line 26 (with an estimated 50 mK rms noise level 
uncertainty) and the assumption that it can be assigned to glycine.

Examination of Table 3 shows that the K$_{-1}$ = 1 or 2 frequencies of the 
degenerate glycine quartets J = 43 - 42, 38 - 37, and 37 - 36 are obscured 
by the wings of lines previously identified by Nummelin et al. (1998).  
U235.085 at 235,085 MHz would be a possible candidate for the J = 39 - 38, 
K$_{-1}$ = 1 or 2, degenerate glycine transition at 235,084.8 MHz, but 
the peak intensity is low.  Finally, the glycine lines missing 
from the Nummelin et al. (1998) data at the $\sim$ 100~mK level are the 
K$_{-1}$ = 1 or 2 frequencies of the degenerate glycine quartets 
J = 42 - 41 at 252,531.295 MHz and 41 - 40 at 246,716.164 MHz. The data 
for these missing lines and for line 26 at 240,900.647 MHz are shown 
in Figures 5(a), 5(b), and 5(c), adapted from the Sgr~B2(N) data 
and molecular assignments of Nummelin et al. (1998).

Table 4 shows that the K$_{-1}$ = 0 or 1 frequencies of the degenerate 
glycine quartets J = 40 - 39, 39 - 38, and 38 - 37 are obscured by the 
wings of lines previously identified by Nummelin et al.  (1998). In addition, 
the spectral region around the degenerate glycine transition 
J = 44 - 43 at 258,820.597 MHz is dominated by what appears to be narrow, weak
interference spikes in the Nummelin et al. (1998) data for Sgr~B2(N), Sgr~B2(M), 
and Sgr~B2(NW). The K$_{-1}$ = 0 or 1 glycine lines missing from the Nummelin 
et al. (1998) data at the $\sim$ 50~mK level are the J = 43 - 42 degenerate 
quartet at 253,005.255 MHz. Furthermore, the J = 42 - 41 degenerate quartet 
at 247,189.500 MHz and the J = 41 - 40 degenerate quartet at 241,373.340 MHz 
(line 27 of Kuan et al. 2003) are missing at the $\sim$ 100~mK level. The data 
for these missing lines are shown in Figures 5(d), 5(e), and 5(f). 
 
Because of the five missing key quartets summarized in Tables 3 and 4 and shown
in Figure 5, we conclude that glycine has not been detected in an
extended source in Sgr B2(N).  In \S 4.1, we ruled out the detection of 
glycine in a compact source in Sgr B2(N-LMH).  Thus, we conclude that 
the presence of interstellar glycine has not been verified in Sgr B2(N).

\subsection{Alternative Identifications}

Now that we have shown that the carriers of key spectral lines assigned to
glycine by Kuan et al. (2003) can not be verified as interstellar glycine, 
in this section we suggest some logical assignments to other, more common 
species. Our proposed carriers, listed in Table 5, are presented to 
underline the impact of line confusion on the growing and difficult problem 
of identifying new interstellar species. The first and second columns in 
Table 5 list the line number and rest frequency for glycine taken from 
Kuan et al. (2003). We list these older, less accurate glycine frequencies 
because they were the only rest frequencies reported by Kuan et al. (2003) 
for their detected lines. They reported that for most of their observations,
filter bank spectrometers with 1 and 2 MHz resolution were used in parallel 
mode, but filter banks of 1 MHz and 500 kHz resolution were used for all 
observations taken from 2000 December to 2001 March. However, it is not 
always clear which spectral resolution applies to each reported glycine 
line since most of their data were not displayed. Therefore, unless 
otherwise stated, we will assume a minimum uncertainty of 1 MHz on top of 
the 2$\sigma$ computational uncertainty in the reported rest frequency 
of each line.  For convenient comparison, the third column in Table 5 
lists our newer, more accurate glycine frequencies taken from Table 1 of 
this paper. The fourth, fifth, and sixth columns list the possible 
molecular candidates, their quantum numbers, and the rest frequencies 
of their transitions, respectively. In \S 1, we pointed out that one of the peculiarities 
in the glycine detection report of Kuan et al. (2003) was that each of 
their three glycine sources (Orion KL, W51 e1/e2, and Sgr~B2(N-LMH)) was 
observed to contain no more than 13 to 16 of the 27 independent spectral 
lines features and only 3 of the 27 spectral line features were found to 
be common to all three sources: lines 21, 24, and 26 in Table 1. 
Consequently, we will begin with a discussion of these lines.
 
Line 21 was observed with the NRAO 12~m telescope with a 30" beam, and was
misassigned to the four-fold degenerate glycine quartet
35$_{0,35}$-34$_{1,34}$, 35$_{1,35}$-34$_{1,34}$, 35$_{0,35}$-34$_{0,34}$, 
and 35$_{1,35}$-34$_{0,34}$ at 206,468 MHz.  In all 3 sources, line 21 is
anomalously strong relative to all other purported glycine transitions 
that were observed.  This suggests that even in the unlikely event that 
all the other lines reported by Kuan et al. (2003) belong to the same 
species, then line 21 most likely belongs to another molecular species 
altogether.  Recently, vibrationally excited ethyl cyanide (CH$_3$CH$_2$CN) 
has been shown to be a common constituent of Sgr B2(N-LMH) (Mehringer 
et al. 2004). Furthermore, the 23$_{6,18}$ - 22$_{6,17}$ A and E and 
23$_{6,17}$ - 22$_{6,16}$ A and E transitions of vibrationally excited 
ethyl cyanide ($\nu_{\rm_b}$=1) span the range of 206,466.6 to 206,467.3 
MHz (J. C. Pearson, private communication 2003). Of particular interest
are the 23$_{6,18}$ - 22$_{6,17}$ A and 23$_{6,17}$ - 22$_{6,16}$ A 
transitions listed in Table 1 because they are at most only 0.85 MHz 
(1.3 km~s$^{-1}$) away from the assumed glycine search frequency of Kuan 
et al. (2003). Therefore, we suggest that a possible carrier of line 
21 is vibrationally excited ethyl cyanide. If this suggestion is correct, 
the slightly weaker ethyl cyanide 23$_{6,18}$ - 22$_{6,17}$ E and
23$_{6,17}$ - 22$_{6,16}$ E ($\nu_{\rm_b}$=1) transitions at 
206,466.623(8) MHz and 206,466.939(7) MHz, respectively, 
are blended into the low frequency shoulder of line 21 in Sgr~B2(N-LMH), 
Orion KL, and W51 e1/e2 (see figures 1(a), (b), and (c) of Kuan et al. 2003).

Line 24 was misassigned to the nearly four-fold degenerate glycine quartet
36$_{3,33}$-35$_{4,32}$, 36$_{4,33}$-35$_{4,32}$, 36$_{3,33}$-35$_{3,32}$, 
and 36$_{4,33}$-35$_{3,32}$ at an average observed frequency of 228,419.1 MHz. 
Kuan et al. (2003) dismissed the frequency coincidence of the J = 24 - 23, 
K$_{-1}$ = 11 transition doublet of vinyl cyanide (CH$_2$CHCN) as an 
"unlikely" $\nu$=2 state of the molecule.  However, 
these transitions are actually in the out-of-plane bending mode, 
$\nu_{15}$=1, which Nummelin \& Bergman (1999) have shown 
to be the carrier of 45 lines of vinyl cyanide in Sgr~B2(N) (including
the vinyl cyanide doublet which was misassigned as line 24 of glycine). In 
their Table 40, Nummelin et al. (1998) list the rest frequency of the vinyl
cyanide doublet as 228,417.8 MHz, but a new calculation by Lovas (2004) gives
228,418.10(17) MHz. As shown in Table 5, the new vinyl cyanide rest frequency 
is  only $\sim$ 1 MHz away from the search frequency which was used by 
Kuan et al. (2003) for glycine, but it is $\sim$ 2 MHz away from the new 
average glycine rest frequency. 

The third key line, line 26 was misassigned to the four-fold degenerate 
glycine quartet 40$_{1,39}$ - 39$_{2,38}$, 40$_{2,39}$ - 39$_{2,38}$, 
40$_{1,39}$ - 39$_{1,38}$, and 40$_{2,39}$ - 39$_{1,38}$ at an average 
observed frequency of 240,899.6 MHz, whereas the actual glycine frequency 
is 1.1 Mhz higher (see Table 5). The Sgr~B2(N) data for line 26 have been 
published by both Kuan et al. (2003) in their Figure 1d and by 
Nummelin et al. (1998) in their Figure 1. We note that line 26 has the same 
flux ($\sim$ 6 Jy) in the $\sim$ 20$"$ beam of Nummelin et al. (1998) as 
it has in the $\sim$ 26$"$ beam of Kuan et al. (2003). This suggests that 
the carrier of line 26 could also be a constituent of a somewhat compact 
source, such as a hot molecular core.

In addition to lines 21, 24, and 26, there are potential carriers for 
other reported glycine lines that were detected in only one or two
of the three reported glycine sources. Several potentially viable alternative 
identifications were listed by Kuan et al.(2003) as possible line carrier 
candidates, but dismissed without further consideration. These worthy 
candidates will be discussed in order by source and listed in the remaining 
sections of Table 5.

Line 12, reported in Sgr~B2(N-LMH) and W51 e1/e2, was misassigned to the 
degenerate 
glycine pair 22$_{12,11}$ - 21$_{12,10}$ and 22$_{12,10}$ - 21$_{12,9}$ 
at an observed frequency of 150,909.8 MHz. CH$_2$DCH$_2$CN (Table 5) was 
considered and dismissed by Kuan et al. (2003) because of the combination 
of rare isotopic species (D) and an unlikely transition. However, because 
of the close frequency coincidence, it can not be completely eliminated from
consideration without further observation of comparable transitions of 
that species.
 
Line 10, observed only in Sgr~B2(N-LMH), was misassigned to the cluster of
glycine lines 24$_{1,23}$ - 23$_{2,22}$, 24$_{2,23}$ - 23$_{2,22}$,
24$_{1,23}$ - 23$_{1,22}$, 22$_{3,19}$ - 21$_{3,18}$, and 
24$_{2,23}$ - 23$_{1,22}$ at an average center frequency of 147,813.7 MHz
(Table 5). Kuan et al. (2003) recognized the possibility of overlap with the
21$_{4,17}$-20$_{5,16}$ transition of CH$_3$CH$_2$OH (ethanol),
but they misidentified it as a $\nu$ = 1 vibrationally excited state of 
ethanol and hence removed it from consideration as a possible identification.
The correct assignment is to g-CH$_3$CH$_2$OH (gauche ethanol) in 
the ground state (Table 5), so ethanol should remain in consideration as 
a possible carrier of line 10.

Line 16, reported only in Orion KL, was misassigned to the 26$_{2,24}$
 - 25$_{2,23}$ transition of glycine at 164,870.0 MHz (Table 5). This line 
is shown to be barely above the noise level by Kuan et al (2003) in their 
Figure 3. They recognized that the frequency of line 16 agrees almost 
perfectly with the degenerate transitions of (CH$_3$)$_2$CO (acetone) listed
in Table 5, but dismissed this possibility because the transitions are 
unfavorable. However, because of the close frequency agreement and the 
low intensity of line 16, there is no strong basis for ignoring acetone 
without further observations.

Line 13, reported only in W51 e1/e2, was misassigned to the
23$_{6,18}$ - 22$_{6,17}$ transition of glycine at 160,153.458(266) MHz
(Table 5). Line 13 was not reported in either Sgr~B2(N-LMH) or Orion,
but the detection of U160144 was listed instead. No spectra of line 13 
or U160144 were displayed by Kuan et al. (2003). Thus it is fortunate that
Lee \& Cho (2002) used the TRAO 14~m telescope to conduct an Orion survey 
over this frequency range. They detected and identified
the 8$_{1,8}$-7$_{1,7}$ transition of H$_2$C$_2$O (ketene) at 160,142 MHz,
with a high frequency shoulder. An examination of their published spectra 
leaves no doubt that U160144 in Orion is ketene, and the high frequency 
shoulder (which extends to almost 160,160 MHz)
includes the frequency of line 13. A further complication is that
line 13 is almost coincident with transitions of both vinyl cyanide 
(CH$_2$CHCN)
and acetone((CH$_3$)$_2$CO) (see Table 5). The vinyl cyanide line was 
considered and ruled out by Kuan et al. (2003) because it appeared to be 
an unfavorable transition in the $\nu$=2 vibrational state. However, it 
is actually in the $\nu_{15}$=1 state. Thus we have listed several 
possibilities for line 13: it could be a velocity component of ketene; it
could be acetone; it could be vinyl cyanide; or it could be a partial 
mixture of these and unknown components. An observational check on these 
possibilities would be straightforward.

Probably there are other good alternative identifications for the
additional lines reported by Kuan et al. (2003), due to the plethora
of transitions from ground and excited states of many common interstellar 
molecules. For example, deep searches for acetone have only been conducted 
in Sgr~B2(N) (Snyder et al. 2002), but acetone may well be detectable in 
Orion KL and 
W51 e1/e2 at the levels cited by Kuan et al. (2003). Certainly most of the
good alternative candidates for lines reported as glycine are more likely 
to be detectable than glycine, and the identification of these candidates 
can be checked by further observational work on other favorable transitions.

\section{ROTATIONAL TEMPERATURE DIAGRAM LIMITATIONS}

In a given spectral range for a given source, it is our intention to
demonstrate that one can use random U line data that are nearly
frequency coincident with glycine transitions to obtain a reasonable
looking rotational diagram.  The results of the rotation diagram are
then used to show that the most likely glycine transitions (particularly
a-type with large transition strengths) are either severely lacking in
predicted intensity or altogether missing in the given spectral range. 

Kuan et al. (2003) used rotation temperature diagrams constructed from 
Sgr B2(N-LMH), Orion KL, and W51 e1/e2 data to bolster their arguments
for the identification of glycine. In Appendix A, we have noted that
rotational temperature diagrams cannot be used as independent tools 
to verify interstellar spectral line identifications because the frequency 
parameter in equation A4 or A5 will dominate the integrated intensity, 
line strength, and dipole moment terms for typical spectral line data in 
the millimeter or submillimeter range. In the following discussion, we
illustrate this important point. As a data base, we used the 123 unidentified 
(U) lines detected in a millimeter wavelength spectral line survey conducted
by Friedel et al.\ (2004) toward Sgr~B2(N-LMH) with the BIMA Array. 
We searched this U line data base for near frequency-coincident transitions 
of glycine with the following criteria for matching lines:

\begin{enumerate} 
\item The 2$\sigma$ frequency uncertainty of each calculated glycine transition had to be less than 2.82 MHz (equivalent 
to twice the largest 1$\sigma$ uncertainty from Kuan et al. 2003). This 
uncertainty turned out to be less than 1.84 MHz.  
\item Friedel et al.\ (2004) established the rest frequency of each U line by 
assuming a v$_{LSR}$ of 64 km s$^{-1}$. Kuan et al.\ (2003) states that an 
acceptable velocity range for Sgr~B2(N-LMH) is between 58 and 75 km s$^{-1}$. 
Therefore, we allowed a tolerance in our U line frequencies which corresponded
to a v$_{LSR}$ of 64$^{+11}_{-6}$ km s$^{-1}$ with an additional uncertainty 
of $\pm$ 390 kHz due to the spectral resolution used by Friedel et al.\ (2004) 
for the U line observations.
\item The width of each matched U line had to be between 2.8 and 12.7 km s$^{-1}$, 
which is the spread in line widths from the data in Kuan et al.\ (2003) toward 
Sgr~B2(N). There were 2 exceptions, but the 
uncertainties easily reached into this range.
\item We selected only U lines whose intensity was not too great for the associated
transition. If the U line appeared to be a blend of transitions from several species,
it was not used.
\end{enumerate}

\noindent Table 6 lists the glycine and U line frequencies that will be used in our 
rotational temperature diagram discussion. The first two columns list calculated 
glycine rest frequencies and corresponding transitions. The third and fourth columns 
list the product of the line strength and dipole moment squared ($S\mu^2$)
and upper energy ($E_{\rm u}$) for each glycine transition. The 
fifth column lists U lines which were selected using criteria listed above. The sixth 
and seventh columns list the U line integrated intensity derived from Friedel et al.\ (2004) 
and synthesized beam size employed in the observations. Table 6 shows that the 
frequencies of 11 U lines (col.\ 5) nearly match frequencies of 12 degenerate and 
non-degenerate glycine transitions (col.\ 1). Under the assumption of no beam dilution, 
the glycine rotational temperature diagram in Figure 6 was constructed using calculated 
glycine frequencies in equation A4. The Figure 6 data points were fit with a straight 
line by a least squares method that weights each integrated intensity by its uncertainty 
which generally results in smaller uncertainties in slope and intercept when compared 
to an unweighted least squares fit (Bevington \& Robinson 1992). The resulting weighted 
fit\footnotemark[11]\footnotetext[11]{Kuan et al.\ (2003) constructed rotation 
temperature diagrams based on fits that were not weighted by the uncertainties in 
the integrated intensities.} produced a rotational temperature of 90(13) K and a total 
column density, $<N_T>$, of 1.1(1)$\times$10$^{17}$ cm$^{-2}$. The derived rotational 
temperature and column density is similar to those of other molecules detected toward 
SgrB2(N-LMH) at high spatial resolution. If Figure 6 is a valid glycine rotational 
diagram, then the derived rotational temperature and column density parameters can be
used to predict other likely (i.e., large S$\mu^2$) glycine lines in the frequency range 
of the Friedel et al.\ (2004) survey. Five such glycine transitions were contained 
within the survey, but none was detected. The 1 $\sigma$ integrated intensity 
upper limits are given in column 6 of Table 6. The integrated intensity upper limits are 
the product of the rms noise level in units of Jy beam$^{-1}$ and an assumed linewidth 
of 12.7 km s$^{-1}$ (the largest linewidth reported by Kuan et al.\ 2003). Predictions 
for the minimum expected intensity of these 5 lines are included in footnotes to col.\ 6 
and range from 9 to 18 $\sigma$. These missing lines demonstrate that the Figure 6 
rotational diagram purported to be glycine is based on incorrectly identified and 
misassigned lines. 

We have produced a rotational temperature diagram that looks
reasonable and gives reasonable results for unidentified lines assumed
to be glycine, but we have also demonstrated that these rotational
temperature diagram results have no power to predict the intensities of
other likely glycine lines.  We conclude that this particular rotational
diagram is the result of misassigned U lines and cannot be used as
reliable independent evidence to support the interstellar molecular
identification of glycine.  We note that
equation A4 or A5 will be dominated by the frequency parameter and, 
as a result, one can almost
always construct a reasonable looking rotational temperature diagram from any
forest of weak lines spread across a large enough range in energy of a
complicated asymmetric rotor molecule like glycine. 

\section{CONCLUSIONS} 

As we discussed in \S 1, there are several pecularities if the lines 
reported by Kuan et al. (2003) in Sgr B2 (N-LMH), Orion KL, and W51el/e2 
belong to interstellar glycine.  Therefore, we examined their assignments of 
key glycine transitions to the spectral lines listed in Table 1.  We noted 
that their lack of correction for beam dilution amounted to the tacit 
assumption that their glycine sources have an extended source size with 
$\Theta_s$ $\geq$ 45$"$. In Orion KL, we found that none of the 15 glycine 
line assignments could be verified
because the reported intensities of all 15 lines predict that a major glycine
quartet (J=19-18, K$_{-1}$ = 0 or 1) should have been detected in our Orion 
data from the NRAO 12~m telescope (Figures 1 and 2).  So we can conclude 
that the Orion spectral lines reported by Kuan et al. (2003) (column 8, 
Table 1) are not glycine.  Previously, the VLA data of Hollis et al. (2003b)
established upper limits in Orion for compact glycine sources.  Therefore, 
glycine has not been verified in either extended or in compact sources in Orion.  
In W51 el/e2, 16 lines were assigned to glycine.  Fifteen of the 16 line 
intensities lend to the predictions that the J=19-18, K$_{-1}$ = 0 or 1 
glycine quartets should have been detected in our W51 data from the NRAO 
12~m telescope (Figures 3 and 4). Since our spectra show a negative result, 
we can conclude that the reported W51 spectral lines are also not glycine 
(column 9, Table 1).  
In the direction of Sgr~ B2(N-LMH), only 13 lines were assigned to glycine 
but the situation is slightly more complicated by the high spectral line 
density.  However, it can be simplified by recognizing that, based on both 
line strength values and intensity calculations, the two strongest glycine 
lines should be those assigned to lines 26 and 27 in column 7 of Table 1. 
Line 26 appears with a peak intensity of 300~mK in the Sgr~B2(N) band scan 
data of Nummelin et al. (1998), but the important connected lines, 
J=42-41 and 41-40 in the K$_{-1}$ =l or 2 series, are missing altogether 
(Table 3).  Furthermore, line 27 should be stronger than line
26, but it too is missing, along with the connected J=43-42 and 42-41 lines 
in the K$_{-1}$ = 0 or 1 series (Table 4).  All of these key glycine lines 
strongly connect major K$_{-1}$ = 0, 1, or 2 rotational levels, so their 
absence is not possible if the other Sgr B2(N-LMH) glycine lines are 
assigned correctly. Therefore, we conclude that the Sgr B2(N-LMH) spectral 
lines reported by Kuan et
al.  (2003) (column 7, Table 1) are not glycine.  In addition, the VLA data of
Hollis et al. (2003a) established upper limits for compact glycine sources in
Sgr B2(N-LM).  Consequently, glycine has not been verified in Sgr B2(N-LMH)
either in extended or in compact sources.

If the 27 lines reported by Kuan et al.  (2003) in Orion KL, W51el/e2, and 
Sgr~B2(N-LMH) are not glycine, what are they?  Lines 21, 24, and 26 in Table 1 are
the 3 lines reported to be common to all 3 sources.  We have argued that
vibrationally excited ethyl cyanide (CH$_3$CH$_2$CN) is a logical assignment for line
21, and vibrationally excited vinyl cyanide (CH$_2$CHCN) for line 24.  The 
intensity of line 26 increases as the observational beam width narrows, 
so it could also be a constituent of a somewhat compact source.
Four other reported glycine lines could be possibly assigned to more common 
species listed in Table 5.  Beyond this, it is 
difficult to comment on other possible line carrier candidates because 
Kuan et al. (2003) only published sample spectra for 11 of their 27 lines.
Finally, we have demonstrated that rotational temperature diagrams can not by
themselves be used to identify interstellar molecules.

\bigskip \acknowledgements We are indebted to A. Nummelin for providing data 
from Sgr~B2(N). We thank E. B. Churchwell, E. C. Sutton, and W. D. Watson 
for helpful assistance. We appreciate the useful comments from two referees.
J.  M.  H.  gratefully acknowledges research support from H. A. Thronson, 
Technology Director of the NASA Office of Space Science. We acknowledge 
support from the Laboratory for Astronomical Imaging at the University of 
Illinois, and NSF grants AST 99-81363 and AST 02-28953.  The laboratory 
measurements of the glycine spectrum were supported by STCU under contract 
No. 2132.

\bigskip

\bigskip

\centerline{APPENDIX A}

\bigskip

\centerline{SOME COMMENTS ABOUT ROTATIONAL TEMPERATURE DIAGRAMS}

The conventional method for forming a rotational temperature diagram is to
plot the logarithm of normalized column density vs. upper state rotational 
energy level. However, this approach incorporates several assumptions. First, 
it is assumed that each region in the interstellar cloud has uniform physical 
conditions and that the populations of the energy levels can be characterized 
by a Boltzmann distribution. Second, optically thin conditions are assumed
but it may be possible to correct for this assumption later in the analysis.  
Third, if it is not possible to measure the source size, it must be assumed
the source fills the observational beam. So, if the population distribution 
can be described adequately by a single rotational temperature and if all 
the lines are optically thin, the measured fluxes are proportional to the 
column densities in the upper levels of the transitions being observed.  

For autocorrelation (single-element telescope) observations, we can start 
with equation (1) in \S 3. For cross-correlation (interferometric array) 
observations, we can use (see, for example, Snyder et al. 2001)

$$ <N_T> = {2.04~W_{\rm I}~Z~e^{E_{\rm u}/T_{\rm rot}}\over B \theta_{\rm a} 
\theta_{\rm b} S \mu^2 \nu^3} \times~10^{20}~{\rm cm}^{-2}.  \eqno(A1) $$ 

\noindent In equation (A1), $W_{\rm I} = \int I_\nu dv$ in Jy beam$^{-1}$ km s$^{-1}$, 
and $I_\nu$ is the flux density per beam. $Z$ is the rotational partition 
function, B is the beam filling factor,
$\theta_{\rm a}$ and $\theta_{\rm b}$ are the FWHM synthesized 
beam dimensions in arcsec, S is the linestrength, $\mu^2$ the square of the
dipole moment in Debye$^2$, and $\nu$ is the frequency in GHz. The 
beam-averaged upper level column density, $<N_u>$, is defined by

$$ \frac{<N_u>}{g_u} = \frac{<N_{T}>}{Z}e^{-\frac{E_u}{T_{rot}}}, \eqno(A2) $$ 

\noindent where g$_{u}$ is the statistical weight of the upper level (2J+1).
A standard approach is to take the Naperian (base e) logarithm of equation (A2),

$$ \ln\left(\frac{<N_u>}{g_u}\right)= -\frac{E_u}{T_{rot}} + ln\left(\frac{<N_T>}{Z}\right). \eqno(A3) $$

Then for cross-correlation data, equations (A1) and (A3) give 

$$ \ln\left(\frac{<N_u>}{g_u}\right) = \ln\left(\frac{2.04~W_{\rm I}10^{20}}{B\theta_{\rm a}\theta_{\rm b}S\mu^2\nu^3} \right)
= -\frac{E_u}{T_{rot}}+\ln\left(\frac{<N_{T}>}{Z}\right). \eqno(A4) $$

\noindent A linear least-squares fit to equation (A4) can be used to
construct a rotational
temperature diagram where the slope gives the negative reciprocal of the 
rotational temperature. This temperature can be used 
to estimate the total molecular column density $<N_T>$. The rotational 
temperature will be equal to the kinetic temperature only at sufficiently 
high densities.  A similar
approach can be followed with equations (1) and (A2) to construct a rotational 
temperature diagram from autocorrelation (single-element telescope) data:

$$ \ln\left(\frac{<N_u>}{g_u}\right) = \ln\left(\frac{1.67~W_{\rm T}10^{14}}{S\mu^2\nu}\right) 
= -\frac{E_u}{T_{rot}}+\ln\left(\frac{<N_{T}>}{Z}\right).\eqno(A5) $$

There are two more important caveats about rotational temperature diagrams 
that must be noted. First, if common (Briggsian) base 10 logarithms are used,
the result is

$$ \log_{10}\left(\frac{<N_u>}{g_u}\right)= -\frac{E_u~log_{10}e}{T_{rot}} + \log_{10}\left(\frac{<N_T>}{Z}\right). \eqno(A6) $$

\noindent This approach slightly complicates the rotational temperature term. 
However, it makes the data scatter appear to be smaller since 
$log_{10}\frac{<N_u>}{g_u}$ = 0.434 $ln_e\frac{<N_u>}{g_u}$. The second 
important caveat is in either equations (A4) or equation (A5), the frequency 
parameter will dominate the integrated intensity, line strength, and dipole moment 
terms for typical spectral line data in the millimeter or submillimeter 
range. Hence, rotational temperature diagrams can not be used as independent 
tools to verify interstellar spectral line identifications. This point is 
illustrated in \S 5.

If the emission is not optically thin, the finite opacity 
produces an underestimate in the upper level column density of the observed 
transition and it can also produce an error in the rotational temperature.
Based on the discussion presented by Goldsmith \& Langer (1999),
the optical depth, $\tau$, can be used to define an optical depth
correction factor, $C_\tau$, where

$$C_\tau = \frac{\tau}{1-e^{-\tau}} = \frac{<N_{\rm u}>}
{<N_{\rm u}^{\rm thin}>}.  \eqno(A7)$$

{\noindent}From equations (A1) and (A2), we find for cross-correlation data

$$<N_{\rm u}^{\rm thin}> = \left(\frac{2.04 g_{\rm u}W_{\rm I}\times10^{20}}
{B\theta_{\rm a}\theta_{\rm b}S\mu^2 \nu^3}\right) cm^{-2}.  \eqno(A8) $$

{\noindent}Since $C_\tau \ge 1$, it has been described by Goldsmith \& Langer
(1999) as the factor by which $<N_{\rm u}^{\rm thin}>$ (derived directly from 
$W_{\rm I}$) can appear to be too small because of a finite optical depth. 
Note that $C_\tau$ can be different for each independent transition because 
$\tau$ is proportional to $S\mu^2$. For finite optical depths, we find

$$\ln\left(\frac{<N_{\rm u}^{\rm thin}>}{g_{\rm u}}\right) = 
\ln\left(\frac{2.04 W_{\rm I}\times10^{20}}
{B\theta_{\rm a}\theta_{\rm b}S\mu^2 \nu^3}\right) = 
\ln\left(\frac{<N_T>}{Z}\right) - \ln C_\tau - \frac{E_{\rm u}}{T_{\rm rot}},
\eqno(A9)$$
{\noindent}which is analogous to equation (24) of Goldsmith \& Langer 
(1999). Clearly, $\ln C_\tau$ lowers the ordinate intensity from the 
optically thin value for each point on the rotational intensity 
diagram. We can't calculate a realistic value of $\ln C_\tau$ for glycine 
as no interstellar glycine line identification has been confirmed. However, 
Goldsmith \& Langer (1999) modeled $\tau$ values for methanol (an esablished 
interstellar asymmetric rotor) that ranged from 0.25 to 30. If $\tau = 30$ 
for a given glycine line, that point on the rotational temperature diagram 
would be depressed by 3.4, a significant amount. On the other hand, 
$\tau = 1$ would reduce the ordinate intensity by only 0.5.

\clearpage

\begin{figure}
\plotone{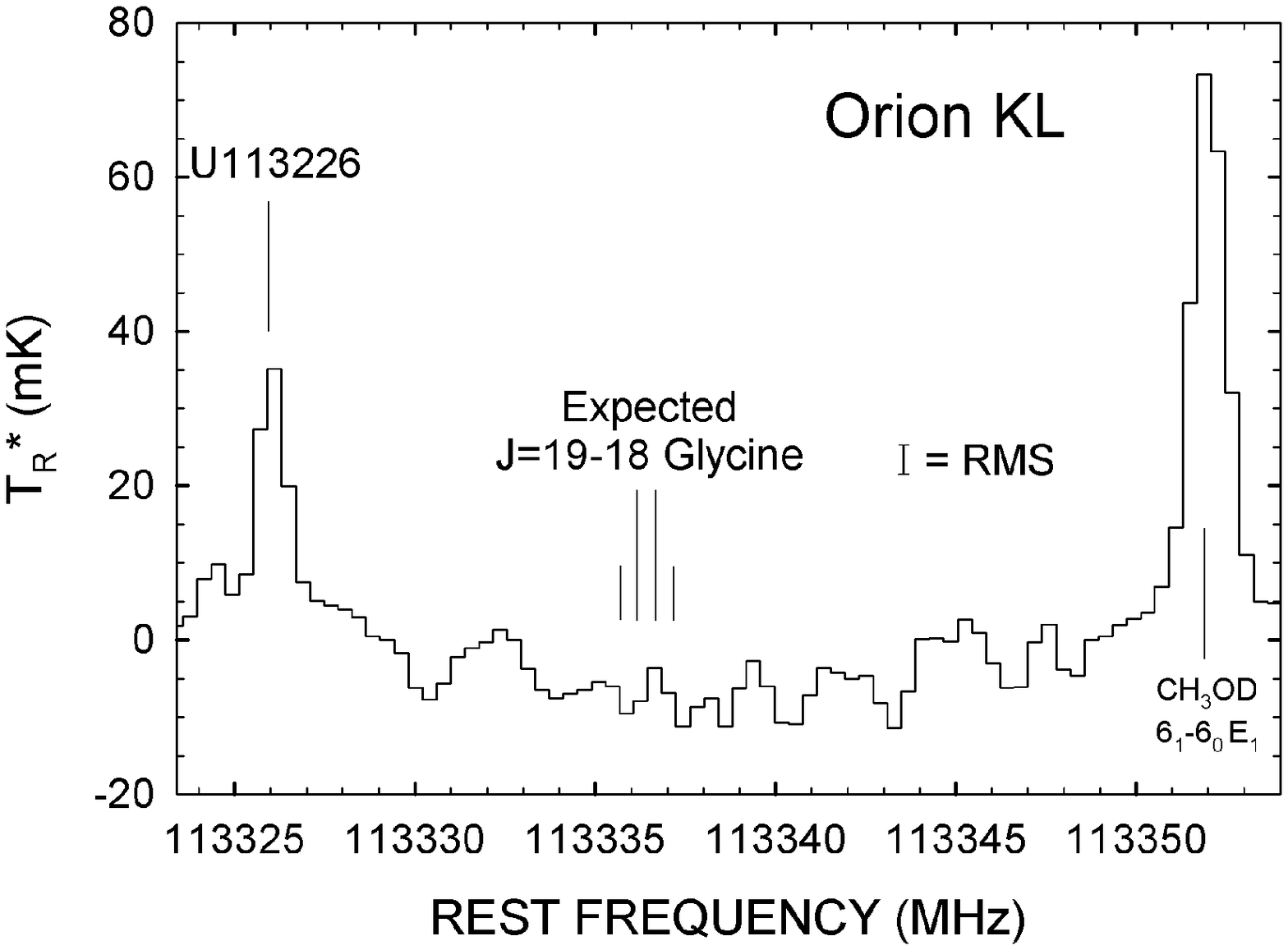}
\caption{Orion KL spectra from 113,323 to 113,355 MHz (centered at
113,339 MHz) observed with the Hybrid Spectrometer on the NRAO 12~m.  The
spectral positions of the negative results for the nearly four-fold 
degenerate J = 19-18 glycine lines are marked by the four vertical lines 
centered at 113,336 MHz.  U113.226 is on the left, and CH$_3$OD is on the 
right.  The ordinate is in units of mK on the T$_R^*$ scale.  The abscissa is rest 
frequency calculated with respect to V$_{\rm LSR}$ = 9 km~s$^{-1}$ (for the 
Orion compact ridge) except the CH$_3$OD rest frequency is with respect
to V$_{\rm LSR}$ = 5.6 km~s$^{-1}$ (representative of the Orion hot core). 
The rms noise level for the spectral region between U113226 and CH$_3$OD 
is 3.7 mK..}
\end{figure}
\clearpage

\begin{figure}
\plotone{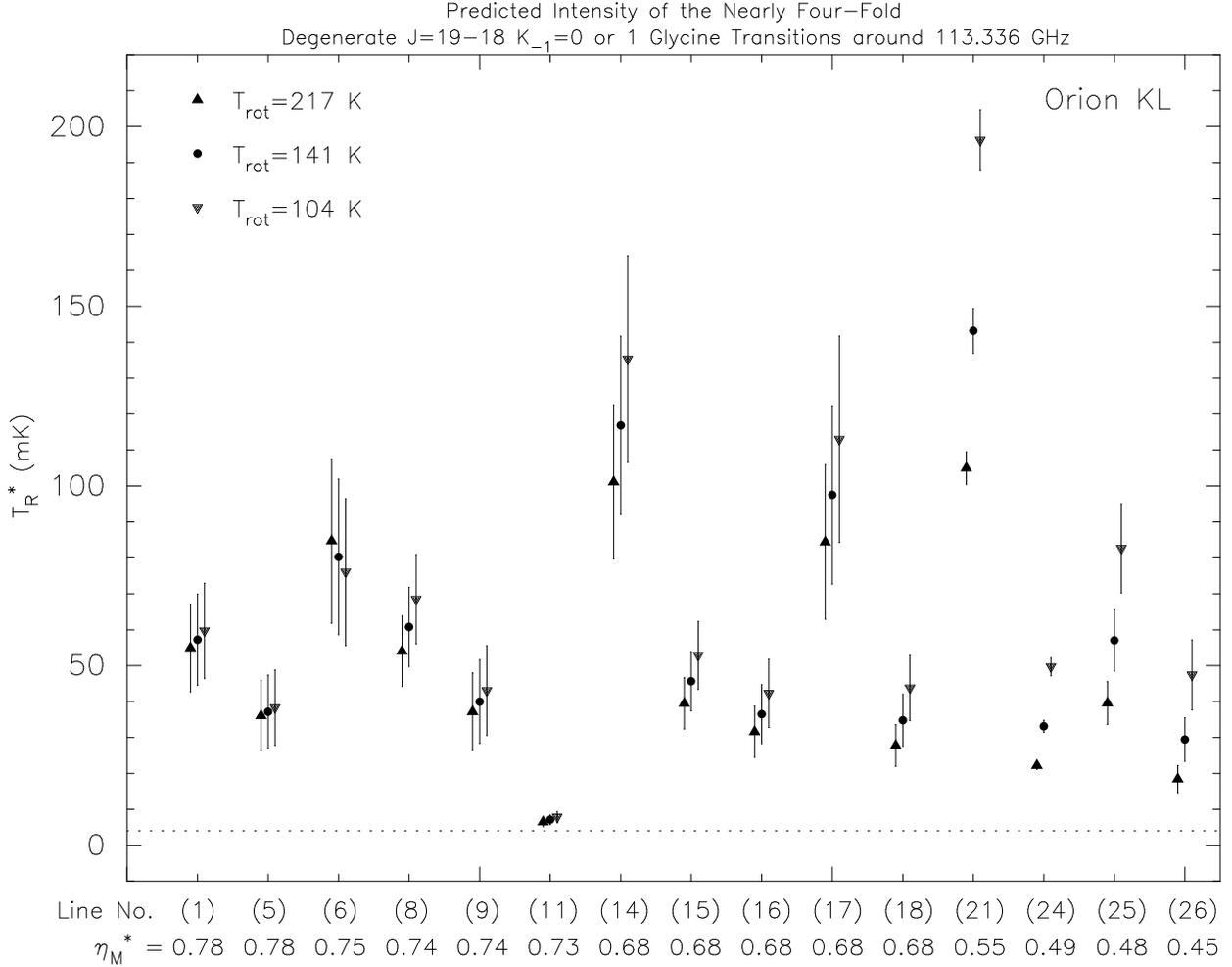}
\epsscale{1.0}
\caption{The intensity of each of the 15 Orion KL glycine lines
reported by Kuan et al. (2003) was used to independently predict the intensity
of the nearly four-fold degenerate J = 19-18 glycine lines at 113.366 GHz. 
A 45$"$ extended source of glycine was assumed.
As in Figure 1, the ordinate is in units of mK based on the T$_R^*$ scale.
The upper abscissa lists the glycine line number from Table 1, and the 
lower gives the corresponding main beam efficiency, $\eta _M^*$,
for the NRAO 12~m. Each predicted intensity is calculated for the Orion 
rotational temperature and its uncertainties, 
T$_{\rm rot}$ = 141$^{+76}_{-37}$ K, derived by 
Kuan et al. (2003) ($\blacktriangle$ = 217 K; $\bullet$ = 141 K; 
$\blacktriangledown$ = 104 K). The uncertainties for each rotational 
temperature point are based on the rms noise level given for each glycine line 
reported by Kuan et al. (2003) (see Table 1). The dotted line shows the 
3.7 mK noise level from Figure 1. The intensities predicted from all 15 
reported glycine lines show that the 19-18 glycine lines should have been 
detected in Orion KL, but they were not (see Figure 1).}
\end{figure}
\clearpage

\begin{figure}
\plotone{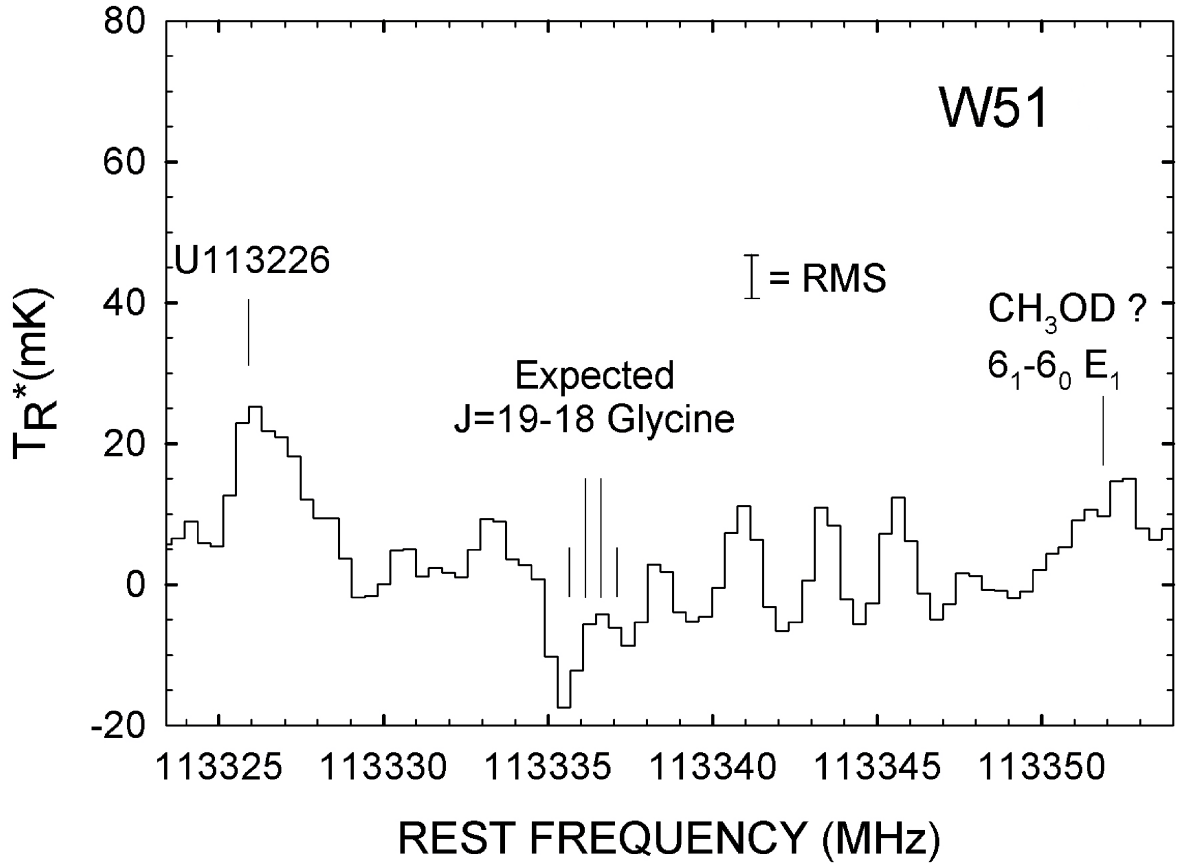}
\caption{W51 spectra from 113,323 to 113,355 MHz (centered at
113,339 MHz) observed with the Hybrid Spectrometer on the NRAO 12~m.  The
spectral positions of the negative results for the nearly four-fold degenerate 
J = 19-18 glycine lines are marked by the four vertical lines centered at 
113,336 MHz.  U113.226 is on the left, and a weak emission feature from
CH$_3$OD is on the right. As in Figure 1, the ordinate is in units of mK on 
the T$_R^*$ scale. The abscissa is rest frequency calculated with respect 
to V$_{\rm LSR}$ = 57.1 km~s$^{-1}$. The rms noise level for the spectral 
region between U113226 and CH$_3$OD is 6.4 mK..}
\end{figure}
\clearpage

\begin{figure}
\plotone{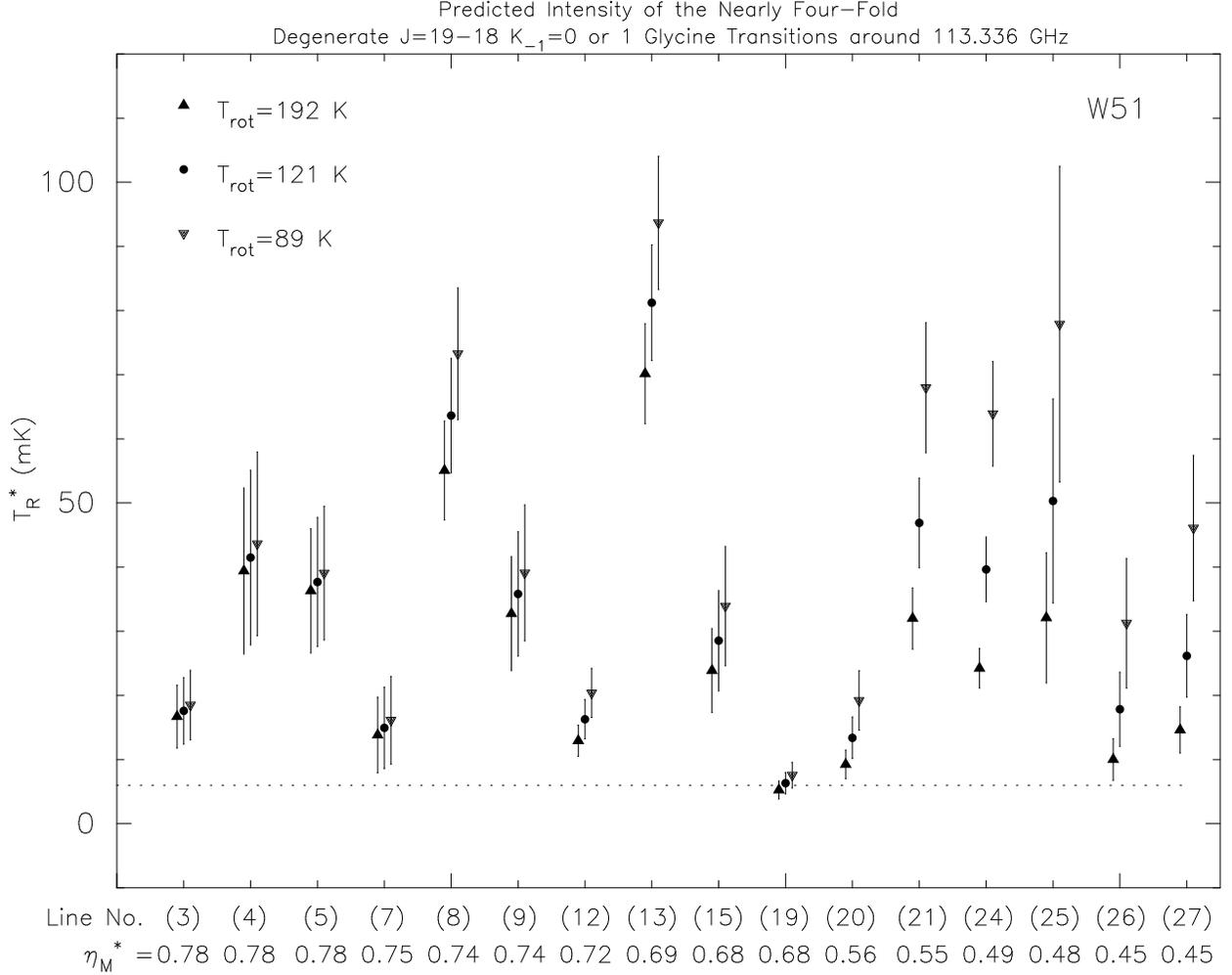}
\epsscale{1.0}
\caption{The intensity of each of the 16 W51 glycine lines
reported by Kuan et al.  (2003) was used to independently predict the intensity
of the nearly four-fold degenerate J = 19-18 glycine lines at 113.366 GHz.  A
45$"$ extended source of glycine was assumed. 
As in Figure 1, the ordinate is in units of mK based on the T$_R^*$ scale.
The upper abscissa lists the glycine line number from Table 1, and the 
lower gives the corresponding main beam efficiency, $\eta _M^*$,
for the NRAO 12~m. Each predicted intensity is calculated for the W51
rotational temperature and its uncertainties, T$_{\rm rot}$ = 
121$^{+71}_{-32}$ K, derived by 
Kuan et al. (2003) ($\blacktriangle$ = 192 K; $\bullet$ = 121 K; 
$\blacktriangledown$ = 89 K). The uncertainties for each rotational 
temperature point are based on the rms noise level given for each glycine line 
reported by Kuan et al. (2003) (see Table 1). The dotted line shows the 
6.4 mK noise level from Figure 3. The intensities predicted from 15 of the 16
reported glycine lines show that the 19-18 glycine lines should have been 
detected in W51, but they were not (see Figure 3).}
\end{figure}
\clearpage

\begin{figure}
\epsscale{0.9}
\plotone{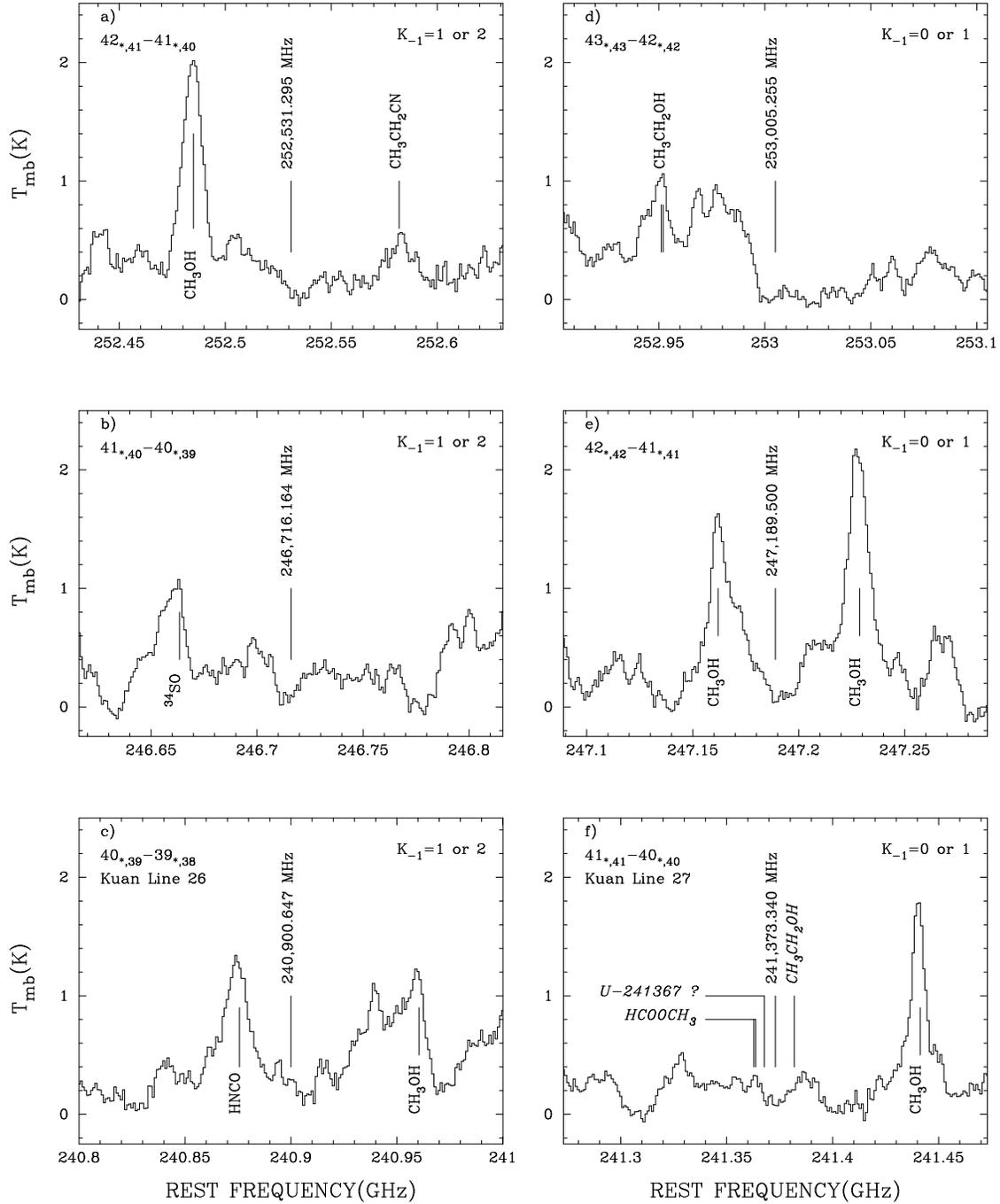}
\epsscale{1.0}
\caption{This figure is adapted from the Sgr~B2(N) data and 
molecular assignments given by Nummelin et al. (1998).
(a) The Nummelin et al. (1998) data for the missing glycine lines for 
the K$_{-1}$ = 1 or 2 frequencies of the fourfold degenerate glycine quartet 
J = 42 - 41 at 252,531.295 MHz.
(b) The missing fourfold degenerate glycine quartet J = 41 - 40 at 246,716.164 MHz.
(c) The data for line 26 of Kuan et al. (2003), which is reported to be the 
fourfold degenerate glycine quartet J = 40 - 39 at 240,900.647 MHz.
(d) The data for the missing glycine lines for the K$_{-1}$ = 0 or 1 frequencies 
of the fourfold degenerate glycine quartet J = 43 - 42 at 253,005.255 MHz.
(e) The missing fourfold degenerate glycine quartet J = 42 - 41 at 247,189.500 MHz.
(f) The missing fourfold degenerate glycine quartet J = 41 - 40 at 241,373.340 MHz 
(line 27 of Kuan et al. 2003).}
\end{figure}
\clearpage

\begin{figure}
\plotone{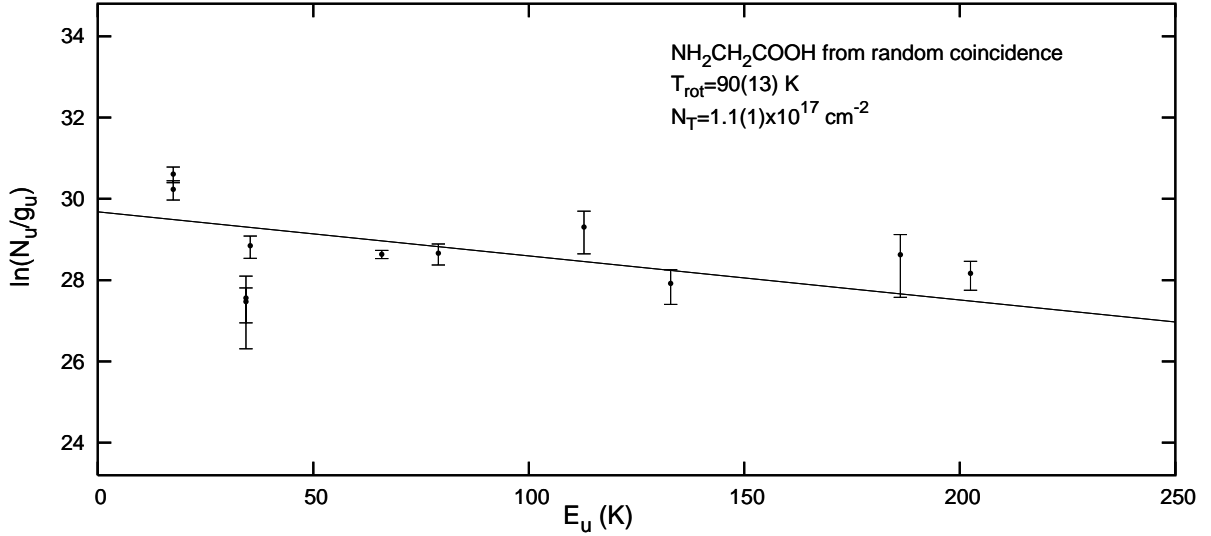}
\epsscale{1.0}
\caption{Pseudo glycine rotational temperature diagram based on 
U lines incorrectly assigned to glycine transitions. Table 6 U line data
(Friedel et al. 2004) were taken with the BIMA array toward SgrB2(N-LMH)
and are approximately frequency coincident with glycine transitions.
The ordinate is the Naperian logarithm of the beam-averaged upper level 
column density, $<N_u>$, divided by g$_{u}$, the statistical weight of 
the upper level (2J+1); the abscissa E$_u$ is the upper state rotational
energy level for each transition. As discussed in the text, this diagram 
may appear to be reasonable and yield a rotation temperature
in a range characteristic of the source, but it has no power to predict
intensities of glycine lines as shown in Table 6.  Such a diagram 
cannot be used as evidence of interstellar glycine (see text). }
\end{figure}

\clearpage

\begin{deluxetable}{l@{}c@{}c@{}c@{}c@{}c@{}c@{}c@{}c@{}c@{}c@{}c@{}c@{}c}
\tabletypesize{\tiny}
\tablewidth{0pt}
\tablecaption{Summary of Reported Glycine Line Detections}
\tablehead{\colhead{} & \colhead{} & \colhead{} & \colhead{Rest} & \colhead{S} & \colhead{} & \multicolumn{2}{c}{Sgr B2(N-LMH)} & \colhead{} & \multicolumn{2}{c}{Orion-KL} & \colhead{} & \multicolumn{2}{c}{W51 e1/e2}\\
\cline{7-8} \cline{10-11} \cline{13-14} \colhead{Line} & \colhead{} & \colhead{} & \colhead{Frequency} & \colhead{Line} & \colhead{E$_u$} & \colhead{v$_{LSR}$} & \colhead{T$_R^*$} & \colhead{} & \colhead{v$_{LSR}$} & \colhead{T$_R^*$} & \colhead{} & \colhead{v$_{LSR}$} & \colhead{T$_R^*$}\\
\colhead{No.} & \colhead{Transition} & \colhead{Type} & \colhead{(MHz)\tablenotemark{a}} & \colhead{Strength} & \colhead{(K)} & \colhead{(km s$^{-1}$)\tablenotemark{b}} & \colhead{(mK)\tablenotemark{c}} & \colhead{} & \colhead{(km s$^{-1}$)\tablenotemark{b}} & \colhead{(mK)\tablenotemark{c}} 
& \colhead{} & \colhead{(km s$^{-1}$)\tablenotemark{b}} & \colhead{(mK)\tablenotemark{c}}}
\startdata
1 & $21_{1,20} - 20_{2,19}$ & b & 130346.775(13) & 16.7 & 72.3 & 66.3(1.2) & 26.0(5.9) && 8.0(3.1) & 9.0(2.0) && \multicolumn{2}{c}{U Interloper\tablenotemark{d}} \\
2 & $21_{2,20} - 20_{2,19}$ & a & 130354.371(13) & 20.5 & 72.3 & 66.3(1.2) & 30.0(5.9) && \multicolumn{2}{c}{U Interloper} & & \multicolumn{2}{c}{U Interloper} \\
3 & $21_{1,20} - 20_{1,19}$ & a & 130360.890(13) & 20.5 & 72.3 & \multicolumn{2}{c}{U Interloper} && \multicolumn{2}{c}{U Interloper} && 60.0(3.1) & 7.5(2.2) \\
4 & $21_{2,20} - 20_{1,19}$ & b & 130368.485(13) & 16.7 & 72.3 & \multicolumn{2}{c}{ID-Interloper\tablenotemark{e}} && \multicolumn{2}{c}{ID-Interloper} && 60.0(3.1) & 6.4(2.1) \\
5 & $19_{3,16} - 18_{3,15}$ & a & 131423.627(13) & 18.1 & 67.6 & \multicolumn{2}{c}{Not Observed\tablenotemark{f}} && {\bf 8.5(1.1)} & {\bf 15.0(4.1)} && 60.0(3.0) & 15.0(4.0) \\
6 & $12_{5,8} - 11_{4,7}$ & b & 142225.340(120) & 4.6 & 34.0 & 64.0(5.6) & 22.0(2.0) && 7.8(2.8) & 5.2(1.4) && \multicolumn{2}{c}{Not Detected\tablenotemark{g}} \\
7 & $21_{3,18} - 20_{3,17}$ & a & 142301.054(13) & 20 & 81.0 & 64.0(2.1) & 38.0(2.2) && \multicolumn{2}{c}{ID-Interloper} && 60.0(2.8) & 6.6(2.8) \\
8  & $21_{9,12} - 20_{9,11}$ & a & 144841.736(22) & 17.2 & 102.9 & \multicolumn{2}{c}{Not Observed} && 5.0(2.8) & 20.9(3.8) && 56.0(2.8) & 20.7(2.9) \\
9 & $21_{5,17} - 20_{5,16}$ & a & 144847.191(16) & 19.7 & 84.8 & \multicolumn{2}{c}{Not Observed} && 8.0(1.4) & 17.9(5.2) && 60.0(1.4) & 15.5(4.2) \\
10 & $24_{1,23} - 23_{2,22}$ & b & 147812.030(17) & 19.7 & 92.7 & 64.0(1.0) & 20.0(5.0) && \multicolumn{2}{c}{Not Observed} && \multicolumn{2}{c}{Not Observed} \\
   & $24_{2,23} - 23_{2,22}$ & a & 147813.168(17) & 23.5 & " && &&  & &&  & \\
   & $24_{1,23} - 23_{1,22}$ & a & 147814.186(17) & " & " && &&  & &&  & \\
   & $22_{3,19} - 21_{3,18}$ & a & 147815.031(14) & 21 & 88.0 && &&  & &&  & \\
   & $24_{2,23} - 23_{1,22}$ & b & 147815.324(17) & 19.7 & 92.7 && &&  & &&  & \\
11 & $25_{0,25} - 24_{1,24}$ & b & 148268.007(18) & 23.6 & 94.0 & \multicolumn{2}{c}{U Interloper} && 5.0(2.0) & 11.0(2.1) && \multicolumn{2}{c}{U Interloper} \\
   & $25_{1,25} - 24_{1,24}$ & a & 148268.016(18) & 24.8 & " && &&  & &&  & \\
   & $25_{0,25} - 24_{0,24}$ & a & 148268.024(18) & " & " && &&  & &&  & \\
   & $25_{1,25} - 24_{0,24}$ & b & 148268.033(18) & 23.6 & " && &&  & &&  & \\
12 & $22_{12,11} - 21_{12,10}$ & a & 150909.703(20) & 15.5 & 130.8 & 68.0(2.7) & 18.0(2.5) && \multicolumn{2}{c}{Not Detected(S/N)\tablenotemark{h}} && 58.0(2.0) & 8.0(1.5) \\
   & $22_{12,10} - 21_{12,9}$ & a & " & " & " && &&  & &&  & \\
13 & $23_{6,18} - 22_{6,17}$ & a & 160153.431(30) & 21.4 & 103.5 & \multicolumn{2}{c}{U Interloper} && \multicolumn{2}{c}{U Interloper} && 60.0(2.5) & 37.0(4.1) \\
14 & $26_{2,24} - 25_{3,23}$ & b & 164851.728(30) & 18.8 & 113.7 & \multicolumn{2}{c}{Not Observed} && {\bf 7.5(1.2)} & {\bf 21.2(4.5)} & & \multicolumn{2}{c}{U Interloper} \\
15 & $26_{3,24} - 25_{3,23}$ & a & 164862.039(30) & 25.2 & 113.7 & \multicolumn{2}{c}{Not Observed} && {\bf 7.5(1.2)} & {\bf 25.0(4.5)} && 60.0(1.2) & 14.6(4.0) \\
16 & $26_{2,24} - 25_{2,23}$ & a & 164870.438(30) & 25.2 & 113.7 & \multicolumn{2}{c}{Not Observed} && {\bf 7.5(1.2)} & {\bf 20.0(4.5)} && \multicolumn{2}{c}{Not Detected} \\
17 & $26_{3,24} - 25_{2,23}$ & b & 164880.749(30) & 18.8 & 113.7 & \multicolumn{2}{c}{Not Observed} && {\bf 7.5(1.2)} & {\bf 17.7(4.5)} && \multicolumn{2}{c}{Not Detected(S/N)} \\
18 & $24_{12,13} - 23_{12,12}$ & a & 164886.068(33) & 18.0 & 146.4 & \multicolumn{2}{c}{Not Observed} && 7.5(1.2) & 21.6(4.5) && \multicolumn{2}{c}{Not Detected} \\
   & $24_{12,12} - 23_{12,11}$ & a & 164886.071(33) & " & " && &&  & &&  & \\
19 & $27_{1,26} - 26_{2,25}$ & b & 165270.666(22) & 22.7 & 115.7 & {\bf 64.8(0.9)} & {\bf 60.0(4.6)} && \multicolumn{2}{c}{Not Detected} && {\bf 58.2(2.4)} & {\bf 8.0(2.1)} \\
   & $27_{2,26} - 26_{2,25}$ & a & 165270.829(22) & 26.5 & " && &&  & &&  & \\
   & $27_{1,26} - 26_{1,25}$ & a & 165270.979(22) & " & " && &&  & &&  & \\
   & $27_{2,26} - 26_{1,25}$ & b & 165271.142(22) & 22.7 & " && &&  & &&  & \\
20 & $33_{2,31} - 32_{3,30}$ & b & 205560.595(67) & 25.7  & 176.8 & {\bf 67.0(0.7)} & {\bf 95.0(11.9)} && \multicolumn{2}{c}{Not Detected} && {\bf 58.6(3.9)} & {\bf 17.0(4.1)} \\
   & $33_{3,31} - 32_{3,30}$ & a & 205560.730(67) & 32.2 & " && &&  & &&  & \\
   & $33_{2,31} - 32_{2,30}$ & a & 205560.849(67) & " & " && &&  & &&  & \\
   & $33_{3,31} - 32_{2,30}$ & b & 205560.984(67) & 25.7 & " && &&  & &&  & \\
21 & $35_{0,35} - 34_{1,34}$ & b & 206468.453(17) & 33.6 & 180.5 & {\bf 66.0(1.9)} & {\bf 210.0(8.9)} && {\bf 7.3(0.7)} & {\bf 230.0(10.0)} && {\bf 60.5(0.7)} & {\bf 65.0(9.7)} \\
   & $35_{1,35} - 34_{1,34}$ & a & " & 34.8 & " && &&  & &&  & \\
   & $35_{0,35} - 34_{0,34}$ & a & " & " & " && &&  & &&  & \\
   & $35_{1,35} - 34_{0,34}$ & b & " & 33.6 & " && &&  & &&  & \\
22 & $36_{0,36} - 35_{1,35}$ & b & 212286.829(18) & 34.6 & 190.7 & 66.0(0.7) & 58.0(15.6) && \multicolumn{2}{c}{ID-Interloper} && \multicolumn{2}{c}{ID-Interloper} \\
   & $36_{1,36} - 35_{1,35}$ & a & " & 35.8 & " && &&  & &&  & \\
   & $36_{0,36} - 35_{0,35}$ & a & " & " & " && &&  & &&  & \\
   & $36_{1,36} - 35_{0,35}$ & b & " & 34.6 & " && &&  & &&  & \\
23 & $36_{2,34} - 35_{3,33}$ & b & 223004.375(88) & 28.9 & 208.2 & {\bf 67.3(1.8)} & {\bf 40.0(10.5)} && \multicolumn{2}{c}{Not Detected} && \multicolumn{2}{c}{Not Observed} \\
   & $36_{3,34} - 35_{3,33}$ & a & 223004.395(88) & 35.2 & " && &&  & &&  & \\
   & $36_{2,34} - 35_{2,33}$ & a & 223004.413(88) & " & " && &&  & &&  & \\
   & $36_{3,34} - 35_{2,33}$ & b & 223004.432(88) & 28.9 & " && &&  & &&  & \\
24 & $36_{3,33} - 35_{4,32}$ & b & 228419.435(104) & 25.9 & 216.3 & 65.0(1.3) & 70.0(7.6) && {\bf 4.9(3.5)} & {\bf 40.0(2.0)} && 57.0(1.8) & 40.0(5.1) \\
   & $36_{4,33} - 35_{4,32}$ & a & 228420.029(104) & 34.9 & " && &&  & &&  & \\
   & $36_{3,33} - 35_{3,32}$ & a & 228420.526(104) & " & " && &&  & &&  & \\
   & $36_{4,33} - 35_{3,32}$ & b & 228421.120(104) & 25.9 & " && &&  & &&  & \\
25 & $33_{6,27} - 32_{6,26}$ & a & 230348.993(64) & 31.3 & 202.5 & \multicolumn{2}{c}{ID-Interloper} && 8.0(3.5) & 26.0(3.9) && 58.7(3.5) & 19.0(6.0) \\
26 & $40_{1,39} - 39_{2,38}$ & b & 240900.647(76) & 35.8 & 244.2 & {\bf 63.2(0.6)} & {\bf 94.0(11.0)} && {\bf 5.0(1.7)} & {\bf 35.0(7.2)} && {\bf 57.0(1.7)} & {\bf 17.0(5.5)} \\
   & $40_{2,39} - 39_{2,38}$ & a & " & 39.5 & " && &&  & &&  & \\
   & $40_{1,39} - 39_{1,38}$ & a & " & " & " && &&  & &&  & \\
   & $40_{2,39} - 39_{1,38}$ & b & " & 35.8 & " && &&  & &&  & \\
27 & $41_{0,41} - 40_{1,40}$ & b & 241373.340(32) & 39.6 & 245.9 & \multicolumn{2}{c}{U/ID-Interloper} && \multicolumn{2}{c}{Not Detected} && {\bf 57.0(1.7)} & {\bf 26.0(6.4)} \\
   & $41_{1,41} - 40_{1,40}$ & a & " & 40.8 & " & & &&  & &&  & \\
   & $41_{0,41} - 40_{0,40}$ & a & " & " & " & & &&  & &&  & \\
   & $41_{1,41} - 40_{0,40}$ & b & " & 39.6 & " & & &&  & &&  & \\
\enddata
\tablenotetext{a}{2$\sigma$ frequency uncertainty is in parentheses.}
\tablenotetext{b}{This is the v$_{LSR}$ and uncertainty listed by Kuan et al. (2003).}
\tablenotetext{c}{This is the T$_R^*$ and rms noise level listed by Kuan et al. (2003).}
\tablenotetext{d}{This indicates that there was an unidentified interloper.}
\tablenotetext{e}{This indicates that there was an identified interloper.}
\tablenotetext{f}{This indicates that no observations were made toward this source for this transition.}
\tablenotetext{g}{This indicates that this line toward this source was not detected but should have been (the predicted signal should have been more than 3$\sigma$).}
\tablenotetext{h}{This indicates that this line toward this source was not detected because it's predicted strength should have been below 3$\sigma$.}
\end{deluxetable}

\clearpage

\begin{deluxetable}{lcccccccccc}
\tabletypesize{\tiny}
\tablewidth{0pt}
\tablecaption{113 GHz Search Frequencies for Orion~KL and W51}
\tablehead{\colhead{Rest} & \colhead{} & \colhead{} & \colhead{} & \colhead{S} & \colhead{} & \multicolumn{2}{c}{Orion-KL} & \colhead{} & \multicolumn{2}{c}{W51 e1/e2}\\
\cline{7-8} \cline{10-11} \colhead{Frequency} & \colhead{} & \colhead{} & \colhead{} & \colhead{Line} & \colhead{E$_u$} & \colhead{T$_R^*$} & \colhead{$\Delta$v} & \colhead{} & \colhead{T$_R^*$} & \colhead{$\Delta$v}\\
\colhead{(MHz)\tablenotemark{a}} & \colhead{Molecule} & \colhead{Transition} & \colhead{Type} & \colhead{Strength} & \colhead{(K)} & \colhead{(mK)} & \colhead{(km s$^{-1}$)} & \colhead{} & \colhead{(mK)} & \colhead{(km s$^{-1}$)}}
\startdata
113,326 & U113.326 & \nodata & \nodata & \nodata & \nodata & 36 & 1.6 & & 23 & 2.9\\
113,335.700 (18) & NH$_2$CH$_2$COOH & $19_{0,19}-18_{1,18}$ & b & 17.6 & 55.5 & $<3.7$ & \nodata & & $<6.4$ & \nodata\\
113,336.209 (18) & NH$_2$CH$_2$COOH & $19_{1,19}-18_{1,18}$ & a & 18.8 & 55.5 & $<3.7$ & \nodata & & $<6.4$ & \nodata\\
113,336.696 (18) & NH$_2$CH$_2$COOH & $19_{0,19}-18_{0,18}$ & a & 18.8 & 55.5 & $<3.7$ & \nodata & & $<6.4$ & \nodata\\
113,337.205 (18) & NH$_2$CH$_2$COOH & $19_{1,19}-18_{0,18}$ & b & 17.6 & 55.5 & $<3.7$ & \nodata & & $<6.4$ & \nodata\\
113,250.80 (10)\tablenotemark{b} & CH$_3$OD & $6_1-6_0 E_1$ & b & 7.8\tablenotemark{c} & 54.8 & 74 & 1.4 & & 11 & 3.6\\
\enddata
\tablenotetext{a}{2$\sigma$ uncertainty is in parentheses.}
\tablenotetext{b}{Rest Frequency from Kaushik, Takage, \& Matsumura (1980)}
\tablenotetext{c}{This is the product S$\mu$$^2$ in Debyes$^2$.}
\end{deluxetable}

\clearpage

\begin{deluxetable}{lccccccc}
\tabletypesize{\tiny}
\tablewidth{0pt}
\tablecaption{Connected K$_{-1}$=1 or 2 Glycine Transitions in the Nummelin et al. Survey Range}
\tablehead{\colhead{} & \colhead{} & \colhead{Rest} & \colhead{S} & \colhead{} & \colhead{Predicted} & \colhead{Observed} & \colhead{}\\
\colhead{} & \colhead{} & \colhead{Frequency} & \colhead{Line} & \colhead{E$_u$} & \colhead{T$_{mb}$} & \colhead{T$_{mb}$} & \colhead{}\\
\colhead{Transition} & \colhead{Type} & \colhead{(MHz)\tablenotemark{a}} & \colhead{Strength} & \colhead{(K)} & \colhead{(mK)} & \colhead{(mK)} & \colhead{Comments}}
\startdata
$43_{2,42}-42_{2,41}$ & a & 258,346.029 (96) & 42.5 & 276.8 & \nodata & \nodata & Masked by CH$_3$NH$_2$ at 258,349 MHz\tablenotemark{b}.\\
$43_{2,42}-42_{1,41}$ & b & 258,346.029 (96) & 38.8 & 276.8 & & & \\
$43_{1,42}-42_{2,41}$ & b & 258,346.029 (96) & 38.8 & 276.8 & & & \\
$43_{1,42}-42_{1,41}$ & a & 258,346.029 (96) & 42.5 & 276.8 & & & \\
\hline
$42_{2,41}-41_{2,40}$ & a & 252,531.295 (89) & 41.5 & 268.2 & $233^{+65}_{-55}$ & $<100$ & \\
$42_{1,41}-41_{2,40}$ & b & 252,531.295 (89) & 37.8 & 268.2 & & & \\
$42_{2,41}-41_{1,40}$ & b & 252,531.295 (89) & 37.8 & 268.2 & & & \\
$42_{1,41}-41_{1,40}$ & a & 252,531.295 (89) & 41.5 & 268.2 & & & \\
\hline
$41_{2,40}-40_{2,39}$ & a & 246,716.164 (82) & 40.5 & 256.1 & $265^{+58}_{-53}$ & $<100$ & \\
$41_{2,40}-40_{1,39}$ & b & 246,716.164 (82) & 36.8 & 256.1 & & & \\
$41_{1,40}-40_{2,39}$ & b & 246,716.164 (82) & 36.8 & 256.1 & & & \\
$41_{1,40}-40_{1,39}$ & a & 246,716.164 (82) & 40.5 & 256.1 & & & \\
\hline
$40_{2,39}-39_{2,38}$ & a & 240,900.647 (76) & 39.5 & 244.3 & Std\tablenotemark{c} & 300& Line 26\tablenotemark{d}.\\
$40_{1,39}-39_{2,38}$ & b & 240,900.647 (76) & 35.8 & 244.3 & & & \\
$40_{2,39}-39_{1,38}$ & b & 240,900.647 (76) & 35.8 & 244.3 & & & \\
$40_{1,39}-39_{1,38}$ & a & 240,900.647 (76) & 39.5 & 244.3 & & & \\
\hline
$39_{2,38}-38_{2,37}$ & a & 235,084.758 (70) & 38.5 & 232.7 & $338^{+73}_{-68}$ & 250 & Line U235085 at 235,085 MHz\tablenotemark{b}.\\
$39_{2,38}-38_{1,37}$ & b & 235,084.758 (70) & 34.8 & 232.7 & & & \\
$39_{1,38}-38_{2,37}$ & b & 235,084.758 (70) & 34.8 & 232.7 & & & \\
$39_{1,38}-38_{1,37}$ & a & 235,084.758 (70) & 38.5 & 232.7 & & & \\
\hline
$38_{2,37}-37_{2,36}$ & a & 229,268.511 (65) & 37.5 & 221.4 & \nodata & \nodata & Masked by CH$_3$CH$_2$CN at 229,262 MHz\tablenotemark{b}.\\
$38_{1,37}-37_{2,36}$ & b & 229,268.511 (65) & 33.8 & 221.4 & & & \\
$38_{2,37}-37_{1,36}$ & b & 229,268.511 (65) & 33.8 & 221.4 & & & \\
$38_{1,37}-37_{1,36}$ & a & 229,268.511 (65) & 37.5 & 221.4 & & & \\
\hline
$37_{2,36}-36_{2,35}$ & a & 223,451.918 (60) & 36.5 & 210.4 & \nodata & \nodata & Masked by NH$_2$CHO at 223,449 MHz\tablenotemark{b}.\\
$37_{2,36}-36_{1,35}$ & b & 223,451.918 (60) & 32.8 & 210.4 & & & \\
$37_{1,36}-36_{2,35}$ & b & 223,451.918 (60) & 32.8 & 210.4 & & & \\
$37_{1,36}-36_{1,35}$ & a & 223,451.918 (60) & 36.5 & 210.4 & & & \\
\enddata
\tablenotetext{a}{2$\sigma$ uncertainty is in parentheses.}
\tablenotetext{b}{Nummelin et al. (1998)}

\tablenotetext{c}{The intensity of this line is the standard on which all other
 Table 3 and Table 4 predictions are based.}
\tablenotetext{d}{Kuan et al. (2003) list this as glycine line 26.}
\end{deluxetable}
\clearpage

\begin{deluxetable}{lccccccc}
\tabletypesize{\tiny}
\tablewidth{0pt}
\tablecaption{Connected K$_{-1}$=0 or 1 Glycine Transitions in the Nummelin et al. Survey Range}
\tablehead{\colhead{} & \colhead{} & \colhead{Rest} & \colhead{S} & \colhead{} & \colhead{Predicted} & \colhead{Observed} & \colhead{}\\
\colhead{} & \colhead{} & \colhead{Frequency} & \colhead{Line} & \colhead{E$_u$} & \colhead{T$_{mb}$} & \colhead{T$_{mb}$} & \colhead{}\\
\colhead{Transition} & \colhead{Type} & \colhead{(MHz)\tablenotemark{a}} & \colhead{Strength} & \colhead{(K)} & \colhead{(mK)} & \colhead{(mK)} & \colhead{Comments}}
\startdata
$44_{1,44}-43_{1,43}$ & a & 258,820.597 (46) & 43.8 & 282.3 & $210^{+72}_{-57}$ & $<200$ & Dominated by narrow noise spikes.\tablenotemark{b}\\
$44_{1,44}-43_{0,43}$ & b & 258,820.597 (46) & 42.6 & 282.3 & & & \\
$44_{0,44}-43_{1,43}$ & b & 258,820.597 (46) & 42.6 & 282.3 & & & \\
$44_{0,44}-43_{0,43}$ & a & 258,820.597 (46) & 43.8 & 282.3 & & & \\
\hline
$43_{1,43}-42_{1,42}$ & a & 253,005.255 (41) & 42.8 & 269.9 & $240^{+68}_{-57}$ & $<50$ & \\
$43_{0,43}-42_{1,42}$ & b & 253,005.255 (41) & 41.6 & 269.9 & & & \\
$43_{1,43}-42_{0,42}$ & b & 253,005.255 (41) & 41.6 & 269.9 & & & \\
$43_{0,43}-42_{0,42}$ & a & 253,005.255 (41) & 42.8 & 269.9 & & & \\
\hline
$42_{1,42}-41_{1,41}$ & a & 247,189.500 (36) & 41.8 & 257.7 & $273^{+62}_{-57}$ & $<100$ & \\
$42_{1,42}-41_{0,41}$ & b & 247,189.500 (36) & 40.6 & 257.7 & & & \\
$42_{0,42}-41_{1,41}$ & b & 247,189.500 (36) & 40.6 & 257.7 & & & \\
$42_{0,42}-41_{0,41}$ & a & 247,189.500 (36) & 41.8 & 257.7 & & & \\
\hline
$41_{1,41}-40_{1,40}$ & a & 241,373.340 (32) & 40.8 & 245.9 & $310^{+54}_{-53}$ & $<100$ & Line 27\tablenotemark{c}. Not masked by interlopers\tablenotemark{b}.\\
$41_{0,41}-40_{1,40}$ & b & 241,373.340 (32) & 39.6 & 245.9 & & & \\
$41_{1,41}-40_{0,40}$ & b & 241,373.340 (32) & 39.6 & 245.9 & & & \\
$41_{0,41}-40_{0,40}$ & a & 241,373.340 (32) & 40.8 & 245.9 & & & \\
\hline
$40_{1,40}-39_{1,39}$ & a & 235,556.787 (29) & 39.8 & 234.3 & \nodata & \nodata & Masked by CH$_3$CH$_2$CN at 235,562 MHz\tablenotemark{b}.\\
$40_{1,40}-39_{0,39}$ & b & 235,556.787 (29) & 38.6 & 234.3 & & & \\
$40_{0,40}-39_{1,39}$ & b & 235,556.787 (29) & 38.6 & 234.3 & & & \\
$40_{0,40}-39_{0,39}$ & a & 235,556.787 (29) & 39.8 & 234.3 & & & \\
\hline
$39_{1,39}-38_{1,38}$ & a & 229,739.850 (26) & 38.8 & 223.0 & \nodata & \nodata & Masked by CH$_3$OH at 229,756 MHz\tablenotemark{b}.\\
$39_{0,39}-38_{1,38}$ & b & 229,739.850 (26) & 37.6 & 223.0 & & & \\
$39_{1,39}-38_{0,38}$ & b & 229,739.850 (26) & 37.6 & 223.0 & & & \\
$39_{0,39}-38_{0,38}$ & a & 229,739.850 (26) & 38.8 & 223.0 & & & \\
\hline
$38_{1,38}-37_{1,37}$ & a & 223,922.537 (23) & 37.8 & 211.9 & \nodata & \nodata & Masked by CH$_3$CH$_2$CN at 223,932 MHz\tablenotemark{b}.\\
$38_{1,38}-37_{0,37}$ & b & 223,922.537 (23) & 36.6 & 211.9 & & & \\
$38_{0,38}-37_{1,37}$ & b & 223,922.537 (23) & 36.6 & 211.9 & & & \\
$38_{0,38}-37_{0,37}$ & a & 223,922.537 (23) & 37.8 & 211.9 & & & \\
\enddata
\tablenotetext{a}{2$\sigma$ uncertainty is in parentheses.}
\tablenotetext{b}{Nummelin et al. (1998)}
\tablenotetext{c}{Kuan et al. (2003) list this as glycine line 27.}
\end{deluxetable}

\clearpage

\begin{deluxetable}{lccccc}
\tabletypesize{\tiny}
\tablecolumns{6}
\tablewidth{0pt}
\tablecaption{Other Candidates for Some of the Reported Glycine Lines}
\tablehead{
\colhead{} & \colhead{Search} & \colhead{Glycine} & \colhead{} & 
\colhead{} & \colhead{Carrier Rest}\\
\colhead{Line} & \colhead{Frequency} & \colhead{Frequency} & 
\colhead{Possible} & \colhead{} & \colhead{Frequency}\\
\colhead{No.}  & \colhead{(MHz)\tablenotemark{a}} & 
\colhead{(MHz)\tablenotemark{a}} & \colhead{Carrier} & 
\colhead{Transition} & \colhead{(MHz)\tablenotemark{a}}}
\startdata
\cutinhead{Lines reported in Sgr B2(N-LMH), Orion KL, and W51 e1/e2}
21 & 206,468.023(1694) & 206,468.453(17) & CH$_3$CH$_2$CN & 
23$_{6,18}$-22$_{6,17}A$ $\nu_{\rm_b}$=1 & 206,467.172(7)\tablenotemark{b}\\
21 & 206,468.023(1694) & 206,468.453(17) & CH$_3$CH$_2$CN & 
23$_{6,17}$-22$_{6,16}A$ $\nu_{\rm_b}$=1 & 206,467.307(7)\tablenotemark{b}\\\\

24 & 228,418.243(1100) & 228,419.435(104) &  CH$_2$CHCN & 24$_{11,13/14}$-
23$_{11,12/13} $ $\nu_{15}$=1 & 228,418.10(17)\tablenotemark{c}\\
24 & 228,418.836(1100) & 228,420.029(104) &  &  & \\
24 & 228,419.333(1100) & 228,420.526(104) &  &  &  \\
24 & 228,419.927(1100) & 228,421.120(104) &  &  &  \\\\

26 & 240,899.571(1908) & 240,900.647(76) & Compact Source (see text)&  &  \\\\

\cutinhead{Lines reported in both Sgr B2(N-LMH) and W51 e1/e2}

12 &150,909.783(166)& 150,909.703(20)& CH$_2$DCH$_2$CN & 4$_{3,1}$-3$_{2,2}$ 
& 150,910.45(12)\tablenotemark{d} \\
12 &150,909.784(166)& 150,909.703(20)& & & \\\\

\cutinhead{Lines reported only in Sgr B2(N-LMH)}

10 & 147,811.758(304) &147,812.030(17)& g-CH$_3$CH$_2$OH &
 21$_{4,17}$-20$_{5,16}$ & 147,815.83(2)\tablenotemark{d}\\
10 & 147,812.896(304) &147,813.168(17)& & & \\
10 & 147,813.913(304) &147,814.186(17)& & & \\
10 & 147,814.763(316) &147,815.031(14)& & & \\
10 & 147,815.051(304) &147,815.324(17)& & & \\\\\\\\\\

\cutinhead{Lines reported only in Orion KL}

16 & 164,870.010(394) & 164, 870.438(30)& (CH$_3$)$_2$CO &
 31$_{12/13,19}$-31$_{11/12,20}$ EE & 164,870.14(18)\tablenotemark{e}\\\\

\cutinhead{Lines reported only in W51 e1/e2}

13 & 160,153.458(266) & 160,153.431(30) & H$_2$C$_2$O & 8$_{1,8}$-7$_{1,7}$ 
& 160,142.1\tablenotemark{d,f}\\
13 & 160,153.458(266) & 160,153.431(30) & CH$_2$CHCN 
& 35$_{6,29}$-36$_{5,32}$ $\nu_{15}$=1 & 160,154.8(52)\tablenotemark{g} \\ 
13 & 160,153.458(266) & 160,153.431(30) & (CH$_3$)$_2$CO 
& 21$_{3,18}$-21$_{2,19}$ AE,EA & 160,155.80(10)\tablenotemark{e}\\\\ 

\enddata

\tablenotetext{a}{2$\sigma$ uncertainty is in parentheses.}
\tablenotetext{b}{Rest frequency from J. C. Pearson, private communication (2003).}
\tablenotetext{c}{Rest frequency from Lovas (2004).}
\tablenotetext{d}{Rest frequency from Pickett et al. (1998).}
\tablenotetext{e}{Rest frequency from Groner et al. (2002).}
\tablenotetext{f}{Identified in Orion by Lee \& Cho (2002).}
\tablenotetext{g}{Rest frequency from F. J. Lovas, private communication (2004).}

\end{deluxetable}

\clearpage

\begin{deluxetable}{ccccccc}
\tabletypesize{\scriptsize}
\tablecaption{Incorrectly Assigned Glycine Transitions Used to Create the Rotational Temperature Diagram}
\tablehead{
\colhead{Calculated\tablenotemark{a}} & \colhead{Corresponding} & \colhead{} & \colhead{} & \colhead{Nearly Frequency} & \colhead{U Line\tablenotemark{b}} & \colhead{Beam}\\
\colhead{Glycine Frequency} & \colhead{Glycine} & \colhead{$S\mu^2$} & \colhead{$E_{\rm u}$} & \colhead {Coincident U Line} & \colhead{$\int \Delta I dv$} & \colhead{$\theta_a \times \theta_b$}\\
\colhead{(MHz)} & \colhead{Transition} & \colhead{(Debye$^2$)} & \colhead{(K)} & \colhead{(MHz)} & \colhead{(Jy beam$^{-1}$ km s$^{-1}$)} &\colhead{($''$$\times$$''$)}\\
\colhead{(1)} & \colhead{(2)} & \colhead{(3)} & \colhead{(4)} & \colhead{(5)} & \colhead{(6)} &\colhead{(7)}}
\startdata
86,210.408(367) & $19_{7,13}-19_{6,14} $ & 5.1 & 79.0 & U86207.8 & 7.1(16) & 23.9$\times$6.6\\
86,440.861(1000) & $32_{9,23}-32_{8,24}$ & 10.4 & 202.5 & U86440.2 & 9.5(28) & 24.1$\times$7.0 \\
86,885.889(19) & $12_{3,9}-11_{3,8}$ & 9.4 & 29.4 & ...  & {\bf $<$1.9\tablenotemark{c}} & 25.7$\times$6.5\\
89,735.673(344) & $26_{7,20}-26_{6,21}$ & 7.4 & 132.9 & U89731.9 & 6.8(23) & 30.9$\times$6.3\\
89,831.975(10) & $14_{1,13}-13_{1,12}$ & 11.2 & 34.4 & U89829.6 & 5.0(17) & 23.8$\times$6.2 \\
89,875.703(14) & $13_{5,9}-12_{5,8}$ & 9.2 & 38.3 & ... & {\bf $<$2.4\tablenotemark{d}} & 24.5$\times$6.2\\
90,035.981(16) & $15_{0,15}-14_{1,14}$ & 6.6 & 35.4 & U90033.9 & 12.3(29) & 23.4$\times$6.6 \\
90,043.198(16) & $15_{1,15}-14_{1,14}$ & 12.3 & 35.4 & ... & {\bf $<$1.8\tablenotemark{e}} & 23.4$\times$6.6\\
90,049.762(16) & $15_{0,15}-14_{0,14}$ & 12.3 & 35.4 & ... & {\bf $<$1.8\tablenotemark{e}} & 23.4$\times$6.6\\
90,056.979(16) & $15_{1,15}-14_{0,14}$ & 6.6 & 35.4 & ... & {\bf $<$2.0\tablenotemark{f}} & 23.8$\times$6.6\\
90,303.355(12) & $14_{2,13}-13_{1,12}$ & 4.6 & 34.4 & U90304.8 & 2.6(14) & 27.1$\times$6.2 \\
106,559.495(497) & $24_{6,18}-23_{7,17}$ & 1.6 & 112.8 & U106559.8 & 5.6(22) & 19.4$\times$5.7 \\
106,767.076(788) & $31_{8,24}-30_{9,21}$ & 1.7 & 186.2 & U106763.3 & 3.0(15) & 20.7$\times$5.3 \\
109,960.322(11) & $16_{8,9}-15_{8,8}$ & 10.0 & 65.9 &  & & \\
                &                     &      &      & U109959.6 & 39.1(37) & 22.0$\times$5.0 \\
109,961.001(11) & $16_{8,8}-15_{8,7}$ & 10.0 & 65.9 &  & & \\
110,089.888(121) & $7_{5,3}-6_{4,2}$ & 2.2 & 17.5 & U110085.6 & 28.3(49) & 18.9$\times$5.3\\
110,106.688(121) & $7_{5,2}-6_{4,3}$ & 2.2 & 17.5 & U110104.7 & 19.5(41) & 18.9$\times$5.3 \\
\enddata
\tablenotetext{a}{2$\sigma$ frequency uncertainty in parenthesis.}
\tablenotetext{b}{Table 5 data derived from Friedel et al.\ (2004).  Limits are 1$\sigma$; integrated intensity 
 uncertainties are 1$\sigma$.}
\tablenotetext{c}{T$_{\rm rot}$ diagram predicts at least a 13$\sigma$ glycine line; none detected.}
\tablenotetext{d}{T$_{\rm rot}$ diagram predicts at least a 9$\sigma$ glycine line; none detected.}
\tablenotetext{e}{T$_{\rm rot}$ diagram predicts at least a 18$\sigma$ glycine line; none detected.}
\tablenotetext{f}{T$_{\rm rot}$ diagram predicts at least a 9$\sigma$ glycine line; none detected.}
\end{deluxetable}

\end{document}